\documentclass[twocolumn]{aastex62}
\usepackage{float}
\usepackage{subfigure}
\usepackage{amsmath}

\received{December 4, 2018} 

\accepted{May 21, 2019}
\submitjournal{ApJ}

\shorttitle{A Multi-epoch X-ray Study of the Spiral Galaxy NGC 7331}
\shortauthors{Jin et al.}

\begin{document}
\title{A Multi-epoch X-ray Study of the Spiral Galaxy NGC 7331}

\correspondingauthor{Ruolan Jin}
\email{jin@gapp.nthu.edu.tw}

\author{Ruolan Jin}
\affiliation{Institute of Astronomy, National Tsing Hua University, Hsinchu 30013, Taiwan}

\author{Albert K. H. Kong}
\affiliation{Institute of Astronomy, National Tsing Hua University, Hsinchu 30013, Taiwan}

\begin{abstract}
X-ray point sources in galaxies are dominated by X-ray binaries (XRBs) that are variables or transients and whether their variability would alter the X-ray luminosity functions (XLF) is still in debate. Here we report on NGC 7331 as an example to test this with 7 Chandra observations. Their detection limit is 7$\times10^{37}$ erg s$^{-1}$ in the 0.3 -- 8.0 keV energy range by assuming a power-law spectral model with a photon index of 1.7. We detected 55 X-ray sources. Thirteen of them are variables for which 3 of them are transients and some of the sources possess a bimodal luminosity-hardness ratio feature, which is often observed among X-ray binaries. Nine more ultra-luminous X-ray sources are found comparing to previous studies and 8 are likely to be low-mass or high-mass XRBs. Twenty-one optical counterpart candidates are found based on the Hubble Space Telescope images, but we cannot rule out the possibility of positional coincidence. The spectral analysis of SN 2014C shows a trend of increasing soft photons and decreasing hydrogen column densities as its outer shell expands. We fit the 7 incompleteness-corrected XLFs to both a power-law (PL) and a power-law with an exponential cut-off (PLC) model using Bayesian method, which is the first time used in XLF fitting. The hierarchical PLC model can describe the XLF of NGC 7331 best with a slope of $\sim$ 0.5 and a luminosity cut-off around 8$\times$10$^{38}$ erg s$^{-1}$. This study proves that multi-epoch observations decrease the deviation due to the variable luminous sources in XLFs. 
\end{abstract}
\keywords{galaxies: individual (NGC 7331) -- X-rays: binaries -- X-rays: galaxies --  supernovae: individual (SN 2014C) }
\section{Introduction}\label{sec:intro}

X-ray population study has long been viewed as an important way of probing the various phases of stellar sources as well as the supernovae heated interstellar diffuse plasma. It is also essential in studying the accretion onto supermassive black hole in galactic nuclei and resulting products due to galactic collisions \citep{fab89, mun03}. Previous study of Galactic X-ray sources shows that X-ray binaries (XRBs) are essential parts of X-ray emission of galaxies. Therefore, by studying extragalactic X-ray population, we can also have a better understanding of different X-ray binary populations in different environment \citep{fab06_1}. For example, X-ray Luminosity Function (XLF) constructed based on X-ray point sources of a galaxy can reflect the environment and provides clues of galaxy evolution. Moreover, we can obtain long-term monitoring of X-ray sources if multiple observations are available. After the launch of the Chandra X-ray Observatory, detailed studies of X-ray sources in nearby galaxies become possible with its unprecedented sub-arc-second spatial resolution. 

In this paper, we report our study of the nearby galaxy NGC 7331 based on multiple Chandra observations. The first deep (29.5 ks) Chandra observation of NGC 7331 was  carried out by Chandra in 2001 and a total of 35 X-ray sources down to L$_{X}\sim$ 5$\times$10$^{37}$ erg s$^{-1}$were detected \citep{zez01}. No other Chandra observation has been taken since then until the discovery of the supernova SN 2014C. There were several observations proposed to monitor SN 2014 C. This series of observations not only provides us a much deeper image to probe the fainter X-ray sources but also offers us a great opportunity in understanding the variability of X-ray sources in NGC 7331. However, there has no in-depth analysis of the X-ray sources of NGC 7331 has been carried out apart from SN 2014C so far. 

NGC 7331 is classified as a SA(s)b galaxy with a high inclination angle of 75\degr and PA = 169\degr. At a distance of 14.7 Mpc, which is estimated by Cepheid variables \citep{fre01}, the $D_{25}$ of 9\farcm 78 corresponds to 40.18 kpc. NGC 7331 belongs to the NGC 7331 group with four other members: NGC 7335, NGC 7336, NGC 7337 and NGC 7340 in the constellation Pegasus. Earlier research \citep{ler08} estimated a star formation rate of 2.99 M$_{\sun}$ yr$^{-1}$. According to the latest results of THINGS (The HI Nearby Galaxy Survey), which uses the VLA data, the average radial gas inflow rate outside of r$_{25}$ is -1.03 $\pm$ 0.7 M$_{\sun}$ yr$^{-1}$ \citep{sch16}. The inflow of the gas is usually considered as fueling the star formation of a galaxy.

Three supernovae, SN 1959D \citep{hum59}, SN 2013bu \citep{ita13} and SN 2014C have been discovered in NGC 7331. Intense observations in different wavelength have been launched to study SN 2014C. It was classified as a hydrogen-poor Type Ib SN \citep{kim14} indicating a mass loss before its explosion. Over the course of one year, it evolved into an H$\alpha$ emission prominent type IIn SN implying its interaction with nearby circumstellar shell \citep{mil15, mar17, bie18}. A series of Chandra observations was proposed after the explosion of SN 2014C, which provide multiple observations of NGC 7331 in X-ray. 

In this study, we report overall properties of the detected X-ray point sources (See Section~\ref{sec:xray}) within the D$_{25}$ region of NGC 7331 observed by seven Chandra observations spanning from January 2001 to October 2016 (see Table~\ref{tab:ChandraLog}). These include light curve analysis, hardness ratios, spectral analysis, and XLFs. In order to look for the optical counterparts of the X-ray sources, we also perform optical analysis of the Hubble Space Telescope (HST) observations of NGC 7331 (See Section~\ref{sec:optical}.) The spectral fitting as well as light curve of SN 2014C at different epoch is also carried out in this study. We adopt December 30, 2013 (MJD = 56656) as explosion date \citep{mar17} throughout the paper.

\begin{figure}[ht]
\figurenum{1}
\epsscale{1}
\includegraphics[scale=0.27]{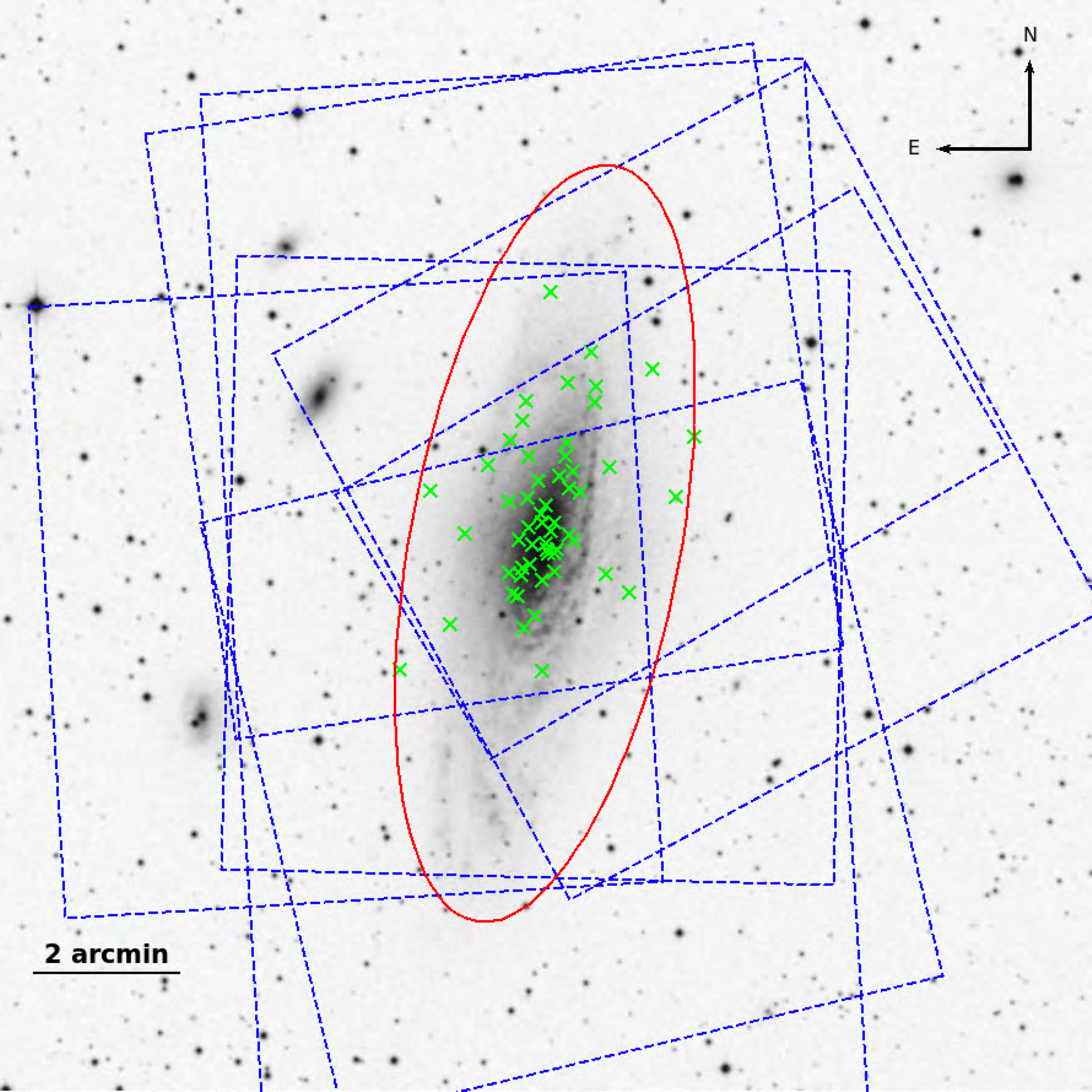}
\caption{The detected X-ray sources (green crosses) overlaid on an optical STScI Digitized Sky Survey NGC 7331 image. The red ellipse is the D$_{25}$ isophote of NGC 7331 and the blue dashed boxes are the coverages of the Chandra ACIS observations used in this study}\label{fig1}
\end{figure}

\section{X-ray observations} \label{sec:xray}

We use seven X-ray observations of NGC 7331 taken by the Advanced CCD Imaging Spectrometer S-array (ACIS-S) of Chandra from 2001 to 2016 in this study (See Table~\ref{tab:ChandraLog} for the observation log). Six of them were observed after SN 2014C explosion. These seven observations cover the entire D$_{25}$ isophote region of NGC 7331 (See Figure~\ref{fig1}). The total exposure time of these seven observations is 126.43 ks. The energy range that we use is between 0.3 and 8.0 keV unless otherwise mentioned. 

\begin{deluxetable}{ccccc}
\tablecaption{Chandra Observation Log \label{tab:ChandraLog}}
\tablecolumns{5}
\tablenum{1}
\tablewidth{\textwidth}
\tabletypesize{\footnotesize}
\tablehead{
\colhead{ObsID} & 
\colhead{Date} & 
\colhead{MJD} & 
\colhead{Exposure} & 
\colhead{Instrument}\\
\colhead{}&
\colhead{(yyyy-mm-dd)}&
\colhead{}&
\colhead{(ks)}&
\colhead{}
}
\startdata
2198 & 2001-01-27 & 51937 & 29.46 & ACIS-S \\
16005 & 2014-11-02 & 56964 & 9.93 & ACIS-S \\
17569 & 2015-01-30 & 57053  & 9.92 & ACIS-S \\
17570 & 2015-04-20 & 57133 & 9.89 & ACIS-S \\
17571 & 2015-08-28 & 57262 & 9.91 & ACIS-S \\
18340 & 2016-05-05 & 57513 & 27.67 & ACIS-S \\
18341 & 2016-10-24  & 57685 & 29.65 & ACIS-S \\
\enddata
\end{deluxetable}

\subsection{Data reduction and source detection\label{subsec:reduction}} 

The data reduction and analysis are done with \texttt{Chandra Interactive Analysis of Observations} (\texttt{CIAO}) version 4.8 (\texttt{CALDB} version 4.7.3), \texttt{HEAsoft} version 6.17 and \texttt{XSPEC} \citep{arn96} version 12.9.0. We reprocess all the Chandra datasets by using \texttt{chandra$\_$repro} script in \texttt{CIAO}. Since six observations (ObsID 16005, 17569, 17570, 17571, 18340, and 18341) were observed in VERY FAINT mode, we also apply the algorithm of ACIS VFAINT background cleaning by enabling the parameter \texttt{check$\_$vf$\_$pha} of the \texttt{chandra$\_$repro} script. 

In order to detect the faint sources, we stack all the seven observations. This step involves reprojecting all the event files to a common tangent point, which means recomputing event sky coordinates. We choose ObsID 2198 as the reference frame and reproject the rest of the other six event lists to match it by using the \texttt{merge$\_$obs} script in \texttt{CIAO}. This script actually runs the \texttt{reproject$\_$obs} to reproject and merges all the stacked files first. Then the exposure maps and the exposure-corrected image are created for the combined image as well for each newly reprojected image by the \texttt{fluximage} script of \texttt{CIAO}. After checking the merged data, we confirm these reprojected images are well aligned and further astrometric correction is not necessary.

We make use of a \texttt{CIAO} implemented Mexican-Hat Wavelet source detection tool \texttt{wavdetect} to detect point sources in the merged and all the reprojected image. \texttt{wavdetect} works by correlating the input image with a series of wavelet scales that matches the size of the sources and are equivalent to the size of the point spread function (PSF). The wavelet are at 9 scales (1, $\sqrt{2}$, 2, 2$\sqrt{2}$, 4, 4$\sqrt{2}$, 8, 8$\sqrt{2}$, and 16 pixels) and we set the detection threshold to be 1$\times 10^{-6}$. Before applying \texttt{wavdetect}, we need to get exposure-corrected images with corresponding exposure maps generated by \texttt{fluximage}. We also need to run the \texttt{mkpsfmap} script to get the PSF map of individual images and the exposure-time weighted PSF map for the combined image. The energy band is set according to 0.3 to 8 keV with an effective energy of 1.5 keV. The output image has been binned to 1 ACIS pixel (=$0\farcs492$) and the central crowded region (the elliptical region with the half radius of optical D$_{25}$ isophote) is set to a subpixel bin size (=1/4 ACIS pixel = $0\farcs123$) instead. 

The detected point sources in all the seven individual images and the merged one are included in the preliminary list. Only those sources lying within the optical $D_{25}$ isophote are considered to be coincident with NGC7331. All the point sources generated by above procedures are further checked visually. The \texttt{wavdetect} routine determined 3$\sigma$ elliptical source region and its nearby source free background region are set for each point source for estimating the count rate. We further adjust the regions by visual inspection to exclude overlapping regions. Only those with source detection significance larger than 3 are included in the final point source candidate list. There are 55 sources detected including the supernova SN 2014C (source No. 42.) We list all the detected sources in Table~\ref{tab:PointSource_table}. All the sources are marked in Figure~\ref{fig1} with crosses.

\begin{figure*}
\figurenum{2}
\epsscale{1}
\plotone{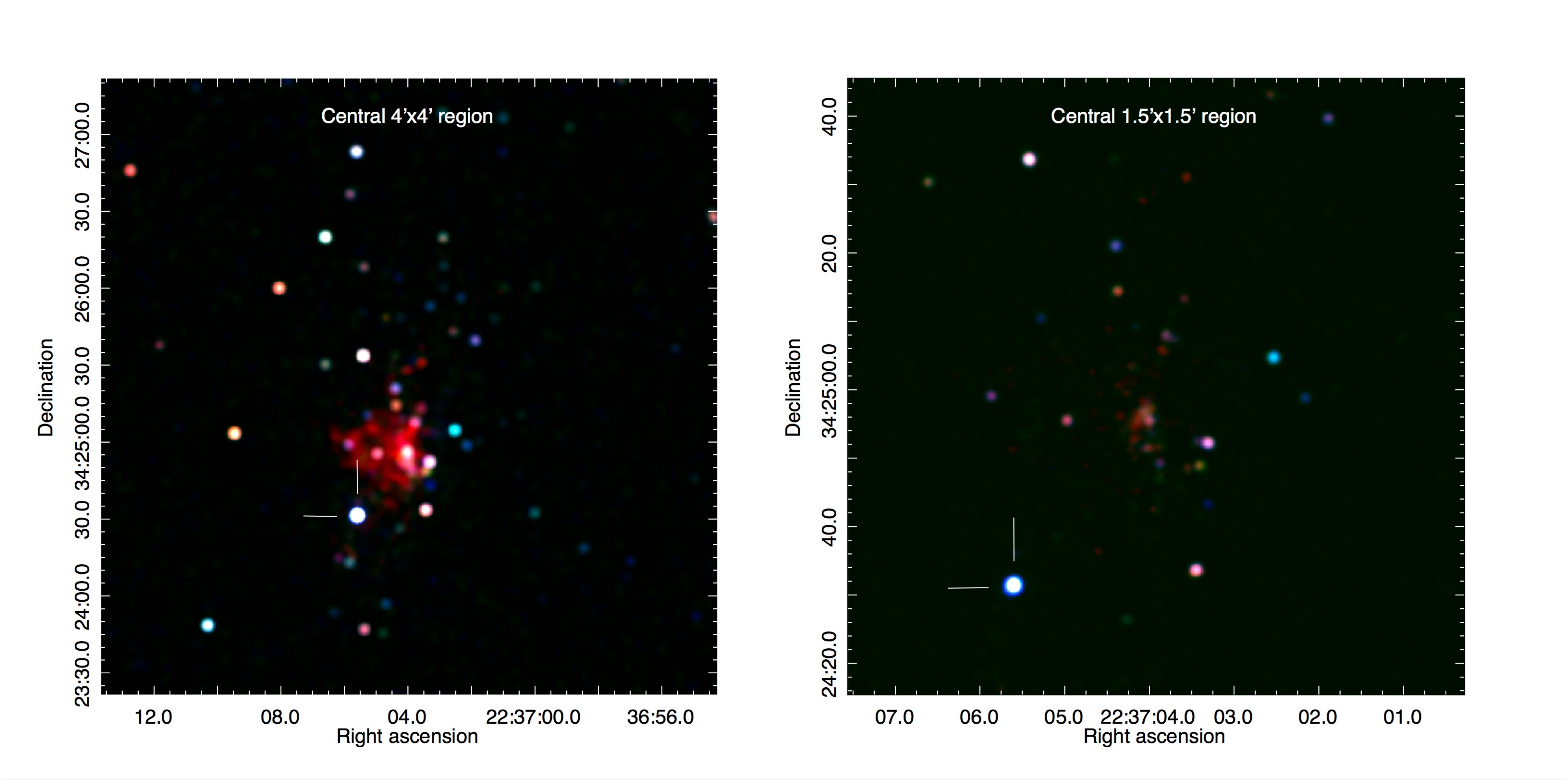}
\caption{The adaptive smoothed three-color Chandra ACIS-S combined images ( ObsID 2198, 16005, 17569, 17570, 17571, 18340, and 18341) of NGC 7331. The RGB channels are red = 0.3 -- 1.0 keV (soft), green = 1.0 -- 2.0 keV (medium) and blue = 2.0 -- 8.0 keV (hard). The left part is the central 4\arcmin$\times$4\arcmin region with pixel size of $0\farcs492$ and the right part is 1.5\arcmin$\times$1.5\arcmin region with pixel size of $0\farcs123$.\label{fig2}}
\end{figure*}

In the central half D$_{25}$ region $\sim$20 kpc), because of the crowded field, we performed source detection on the merged images in both full and subpixel resolution. Figure~\ref{fig2} shows the three-color images of the central region of the stacked Chandra observations of NGC 7331 with different spatial resolutions.

We estimate the expected background X-ray sources based on the Chandra Deep Field North (CDF-N) observations. We use the 2 -- 8 keV X-ray luminosity function of CDF-N \citep{bra01} in the flux range above 1.5$\times 10^{-15}$ erg s$^{-1}$ cm$^{-2}$ to estimate the number of background sources in our field. We converted 0.3 -- 8 keV count rate of the faintest detected source to 2 -- 8 keV flux using WebPIMMS\footnote{https://heasarc.gsfc.nasa.gov/cgi-bin/Tools/w3pimms/w3pimms.pl} by assuming a power-law with a photon index of 1.7 and Galactic absorption. By adopting the converted flux and the sky area in our study to the estimation equation of \cite{bra01}, we estimate $\sim$ 5 background sources in our field.
 
\startlongtable
\begin{deluxetable*}{cccrccrc}
\tablecaption{Chandra Source Properties \label{tab:PointSource_table}}
\tablenum{2}
\tabletypesize{\footnotesize}
\tablewidth{0pt}
\tablehead{ 
\colhead{Source}  & \colhead{RA} & \colhead{DEC} & \colhead{Net\tablenotemark{a}} &\colhead{Count Rate\tablenotemark{b}} & \colhead{Flux\tablenotemark{c}} &\colhead{Lx\tablenotemark{d}} &\colhead{class\tablenotemark{e}}\\ 
\colhead{No.}  \vspace{-0.5cm} & \colhead{(J2000)} & \colhead{(J2000)}  & \colhead{Counts} & \colhead{($\times 10^{-2}$/s)} & \colhead{($\times 10^{-14}$ erg/s/cm$^{2}$)} & \colhead{(erg/s)}& \\
}
\decimals
\startdata
1 & 22:36:54.356 & +34:26:27.20 & 18 & ${0.18(}_{-0.08}^{+0.06})$ & ${2.71(}_{-0.96}^{+1.24})$ & 7.01E+38 & LHa\\
2 & 22:36:54.915 & +34:24:44.90 & 10 & ${0.10(}_{-0.07}^{+0.05})$ & ${1.19(}_{-0.56}^{+0.78})$ & 3.08E+38 &  \\
3 & 22:36:55.513 & +34:25:37.20 & 20 & ${0.07(}_{-0.03}^{+0.03})$ & ${0.93(}_{-0.76}^{+1.05})$ & 2.39E+38 & v\\
4 & 22:36:57.296 & +34:27:21.40 & 9 & ${0.09(}_{-0.07}^{+0.05})$ & ${1.48(}_{-0.54}^{+0.75})$ & 3.82E+38 &  \\
5 & 22:36:58.463 & +34:24:18.10 & 10 & ${0.10(}_{-0.07}^{+0.05})$ & ${1.09(}_{-0.52}^{+0.73})$ & 2.81E+38 &  \\
6 & 22:37:00.018 & +34:24:32.10 & 10 & ${0.10(}_{-0.07}^{+0.05})$ & ${1.13(}_{-0.58}^{+0.84})$ & 2.91E+38 &  \\
7 & 22:37:01.012 & +34:27:06.40 & 9 & ${0.09(}_{-0.06}^{+0.04})$ & ${1.23(}_{-0.27}^{+0.35})$ & 3.17E+38 & B\\
8 & 22:37:01.037 & +34:26:52.80 & 15 & ${0.05(}_{-0.03}^{+0.02})$ & ${0.66(}_{-0.49}^{+0.56})$ & 1.72E+38 &  \\
9 & 22:37:01.348 & +34:27:34.40 & 42 & ${0.14(}_{-0.04}^{+0.04})$ & ${1.86(}_{-0.79}^{+1.00})$ & 4.80E+38 & Ha\\
10 & 22:37:01.872 & +34:25:39.30 & 21 & ${0.21(}_{-0.09}^{+0.07})$ & ${2.38(}_{-0.38}^{+0.46})$ & 6.15E+38 & H/v\\
11 & 22:37:02.156 & +34:24:58.90 & 31 & ${0.11(}_{-0.04}^{+0.03})$ & ${1.35(}_{-0.21}^{+0.26})$ & 3.50E+38 & H\\
12 & 22:37:02.382 & +34:25:56.20 & 19 & ${0.06(}_{-0.03}^{+0.03})$ & ${0.54(}_{-0.65}^{+0.66})$ & 1.39E+38 &  \\
13 & 22:37:02.534 & +34:25:04.20 & 72 & ${0.26(}_{-0.05}^{+0.05})$ & ${3.32(}_{-0.24}^{+0.29})$ & 8.57E+38 & Ha/B\\
14 & 22:37:02.575 & +34:25:42.40 & 22 & ${0.08(}_{-0.03}^{+0.03})$ & ${0.64(}_{-0.26}^{+0.34})$ & 1.64E+38 & a\\
15 & 22:37:02.856 & +34:26:19.80 & 35 & ${0.36(}_{-0.11}^{+0.09})$ & ${4.14(}_{-1.06}^{+1.28})$ & 1.07E+39 & a/v\\
16 & 22:37:02.875 & +34:26:08.20 & 12 & ${0.04(}_{-0.03}^{+0.02})$ & ${0.54(}_{-0.32}^{+0.42})$ & 1.40E+38 &  \\
17 & 22:37:02.916 & +34:27:08.30 & 16 & ${0.06(}_{-0.03}^{+0.02})$ & ${0.87(}_{-0.22}^{+0.30})$ & 2.26E+38 &  \\
18 & 22:37:03.323 & +34:25:52.50 & 11 & ${0.04(}_{-0.02}^{+0.02})$ & ${0.50(}_{-0.75}^{+0.75})$ & 1.30E+38 &  \\
19 & 22:37:03.365 & +34:24:52.20 & 107 & ${0.38(}_{-0.06}^{+0.06})$ & ${4.69(}_{-1.10}^{+1.31})$ & 1.21E+39 & c/Ha\\
20 & 22:37:03.428 & +34:24:33.40 & 37 & ${0.39(}_{-0.12}^{+0.10})$ & ${4.31(}_{-0.70}^{+0.91})$ & 1.12E+39 & a\\
21 & 22:37:03.467 & +34:24:48.30 & 15 & ${0.15(}_{-0.08}^{+0.06})$ & ${1.70(}_{-0.26}^{+0.30})$ & 4.41E+38 & c/a/B\\
22 & 22:37:03.534 & +34:25:13.30 & 22 & ${0.08(}_{-0.04}^{+0.03})$ & ${0.64(}_{-0.82}^{+0.99})$ & 1.66E+38 &  \\
23 & 22:37:03.787 & +34:25:06.50 & 15 & ${0.16(}_{-0.09}^{+0.07})$ & ${1.76(}_{-0.37}^{+0.45})$ & 4.56E+38 & c/L\\
24 & 22:37:03.875 & +34:24:48.30 & 27 & ${0.10(}_{-0.04}^{+0.03})$ & ${1.23(}_{-0.61}^{+0.84})$ & 3.18E+38 & c/LHa\\
25 & 22:37:04.008 & +34:24:51.50 & 10 & ${0.11(}_{-0.07}^{+0.05})$ & ${1.25(}_{-0.34}^{+0.43})$ & 3.23E+38 & ca/B\\
26 & 22:37:04.066 & +34:24:56.20 & 52 & ${0.54(}_{-0.13}^{+0.13})$ & ${6.02(}_{-1.47}^{+1.47})$ & 1.56E+39 & c/La/B\\
27 & 22:37:04.095 & +34:23:11.70 & 26 & ${0.09(}_{-0.03}^{+0.03})$ & ${1.14(}_{-0.21}^{+0.26})$ & 2.96E+38 & L\\
28 & 22:37:04.122 & +34:25:28.20 & 18 & ${0.06(}_{-0.03}^{+0.03})$ & ${0.51(}_{-0.80}^{+1.02})$ & 1.32E+38 &  \\
29 & 22:37:04.154 & +34:28:23.20 & 44 & ${0.17(}_{-0.05}^{+0.04})$ & ${2.63(}_{-0.39}^{+0.47})$ & 6.80E+38 & a/v\\
30 & 22:37:04.252 & +34:24:26.60 & 20 & ${0.21(}_{-0.09}^{+0.07})$ & ${2.37(}_{-0.53}^{+0.53})$ & 6.14E+38 & a\\
31 & 22:37:04.380 & +34:25:13.80 & 29 & ${0.11(}_{-0.04}^{+0.03})$ & ${1.33(}_{-0.22}^{+0.27})$ & 3.45E+38 & La/B\\
32 & 22:37:04.414 & +34:25:20.60 & 48 & ${0.18(}_{-0.04}^{+0.04})$ & ${2.18(}_{-0.28}^{+0.35})$ & 5.64E+38 & Ha/v\\
33 & 22:37:04.697 & +34:23:56.60 & 19 & ${0.06(}_{-0.03}^{+0.02})$ & ${0.80(}_{-0.83}^{+1.02})$ & 2.06E+38 &  \\
34 & 22:37:04.705 & +34:25:47.40 & 22 & ${0.08(}_{-0.03}^{+0.03})$ & ${0.64(}_{-0.58}^{+0.81})$ & 1.65E+38 & a\\
35 & 22:37:04.967 & +34:24:55.20 & 17 & ${0.17(}_{-0.09}^{+0.07})$ & ${1.97(}_{-0.32}^{+0.40})$ & 5.08E+38 & c/a\\
36 & 22:37:05.095 & +34:24:38.70 & 11 & ${0.11(}_{-0.07}^{+0.05})$ & ${1.32(}_{-0.76}^{+0.97})$ & 3.41E+38 &  \\
37 & 22:37:05.274 & +34:25:09.70 & 18 & ${0.06(}_{-0.03}^{+0.03})$ & ${0.79(}_{-0.49}^{+0.72})$ & 2.05E+38 &  \\
38 & 22:37:05.374 & +34:25:33.30 & 142 & ${1.49(}_{-0.21}^{+0.21})$ & ${16.80(}_{-2.34}^{+2.35})\phantom{..}$ & 4.34E+39 & a/vTB\\
39 & 22:37:05.383 & +34:23:46.70 & 20 & ${0.21(}_{-0.09}^{+0.07})$ & ${2.27(}_{-0.21}^{+0.27})$ & 5.87E+38 & LHa\\
40 & 22:37:05.393 & +34:26:07.70 & 9 & ${0.09(}_{-0.06}^{+0.04})$ & ${1.04(}_{-2.80}^{+2.81})$ & 2.69E+38 & L\\
41 & 22:37:05.566 & +34:24:36.30 & 20 & ${0.07(}_{-0.03}^{+0.03})$ & ${0.57(}_{-1.44}^{+1.45})$ & 1.48E+38 &  \\
42 & 22:37:05.614 & +34:24:30.00 & 1437 & ${5.12(}_{-0.23}^{+0.23})$ & ${62.87(}_{-0.28}^{+0.33})\phantom{..}$ & 1.63E+40 & H(SN)/v\\
43 & 22:37:05.628 & +34:26:53.20 & 50 & ${0.53(}_{-0.12}^{+0.12})$ & ${6.19(}_{-0.39}^{+0.47})$ & 1.60E+39 & LH\\
44 & 22:37:05.821 & +34:26:36.50 & 37 & ${0.13(}_{-0.04}^{+0.03})$ & ${1.08(}_{-0.17}^{+0.22})$ & 2.78E+38 & LH\\
45 & 22:37:05.851 & +34:24:13.10 & 49 & ${0.52(}_{-0.12}^{+0.12})$ & ${6.17(}_{-1.12}^{+1.13})$ & 1.60E+39 & a/vT\\
46 & 22:37:05.871 & +34:24:58.90 & 28 & ${0.11(}_{-0.04}^{+0.03})$ & ${1.27(}_{-0.30}^{+0.35})$ & 3.27E+38 & H/B\\
47 & 22:37:06.187 & +34:24:14.30 & 15 & ${0.05(}_{-0.03}^{+0.02})$ & ${0.42(}_{-0.72}^{+0.72})$ & 1.10E+38 &  \\
48 & 22:37:06.424 & +34:24:31.90 & 60 & ${0.21(}_{-0.05}^{+0.05})$ & ${2.60(}_{-1.24}^{+1.44})$ & 6.72E+38 & a/v\\
49 & 22:37:06.590 & +34:26:20.00 & 56 & ${0.21(}_{-0.05}^{+0.05})$ & ${5.13(}_{-1.76}^{+1.77})$ & 1.33E+39 & a/vT\\
50 & 22:37:06.608 & +34:25:30.40 & 38 & ${0.13(}_{-0.04}^{+0.03})$ & ${1.15(}_{-0.56}^{+0.77})$ & 2.99E+38 & L/B\\
51 & 22:37:08.027 & +34:26:00.00 & 195 & ${0.69(}_{-0.08}^{+0.08})$ & ${6.01(}_{-0.27}^{+0.36})$ & 1.55E+39 & a/v\\
52 & 22:37:09.469 & +34:25:02.90 & 47 & ${0.48(}_{-0.13}^{+0.11})$ & ${5.42(}_{-0.62}^{+0.73})$ & 1.40E+39 & La/vB\\
53 & 22:37:10.314 & +34:23:48.20 & 75 & ${0.78(}_{-0.15}^{+0.15})$ & ${9.14(}_{-0.59}^{+0.59})$ & 2.36E+39 & a/v\\
54 & 22:37:11.822 & +34:25:37.50 & 11 & ${0.11(}_{-0.07}^{+0.05})$ & ${1.23(}_{-0.34}^{+0.42})$ & 3.19E+38 &  \\
55 & 22:37:13.514 & +34:23:09.40 & 14 & ${0.05(}_{-0.03}^{+0.02})$ & ${0.65(}_{-1.45}^{+1.46})$ & 1.67E+38 &   
\enddata
\tablenotetext{a}{The net counts are background scaled net counts in the source regions.}
\tablenotetext{b}{The count rates and the uncertainties are obtained by computing the Bayesian background-marginalized posterior probability distribution function (PDF) using the \texttt{aprates} tool.}
\tablenotetext{c}{The 0.3 -- 8 keV maximum unabsorbed model fluxes in these seven observations by assuming a power law model with a photon index of 1.7 and an absorption of Galactic $\mathrm{N_{H}}$, which is evaluated by an interactive program COLDEN. COLDEN is supported by the datasets provided by Bell Laboratories \citep{sta92} and National Radio Astronomy Observatory (NRAO) HI Survey \citep{dic90}.}
 \tablenotetext{d}{Max Lx : The 0.3 -- 8 keV maximum luminosity estimated from the unabsorbed maximum fluxes by assuming a distance of 14.7 Mpc.}
\tablenotetext{e}{The sources classified as variables according to Section~\ref{subsec:flux} are denoted with ``v" and the transients (with maximum to minimum flux $>$ 10) are added with a letter ``T". The variables possess low/hard and high/soft bimodal characters in flux and HR variation are denoted in ``B". ``L", ``H", and ``a" are the classification determined by the X-ray colors adopted from \cite{pre03}; L: Low mass X-ray binary, H: High-mass X-ray binary, a: Absorbed sources. SN: SN 2014C. c: The sources located within less than 1.5" radius of the central region of NGC 7331. }
 \end{deluxetable*}

\subsection{X-ray fluxes and variability} \label{subsec:flux}

The X-ray flux as well as the 68\% confidence intervals of each source is calculated by applying an absorbed power-law model with a photon index ($\Gamma$) of 1.7 with the source and background regions determined in Section~\ref{subsec:reduction} using the \texttt{srcflux} script of \texttt{CIAO} in the seven observations. Taken the PSF contributions of both the source and background regions into account, the net count rate and 90$\%$ confidence interval for point sources are calculated using the \texttt{aprates} tool. The Galactic hydrogen column densities ($\mathrm{N_{H}}$) are obtained from an interactive program \texttt{COLDEN}, which can calculate Galactic $\mathrm{N_{H}}$ according to the datasets supported by the Bell Laboratory \citep{sta92} and National Radio Observatory (NRAO) \citep{dic90}. The conversion factor for converting from count rate to flux in the \texttt{srcflux} tool is using the \texttt{modelflux} script to calculate with the given spectral model. ARFs (Auxiliary Response Files) and RMFs (Redistribution Matrix Files) are also created for each source. By applying the above parameters, the unabsorbed 0.3 -- 8.0 keV energy fluxes as well as the 68$\%$ confidence intervals of each X-ray source of each observation can be estimated. 

Since the flux varies from observation to observation, we list the maximum flux of each source in Table \ref{tab:PointSource_table}. The unabsorbed 0.3 -- 8.0 keV fluxes of the detected sources are in the range of $\sim 10^{-15}$ to $10^{-13}$ erg s$^{-1}$ cm$^{-2}$ corresponding to luminosities of $\sim 10^{38}$ to $10^{40}$ erg s$^{-1}$ with an adopted Cepheid distance of 14.7 $\pm$ 0.6 Mpc to the host galaxy NGC 7331 \citep{fre01}. The maximum luminosities of the detected X-ray point sources range from the faintest 1.10$\times 10^{38}$ erg s$^{-1}$ (source No. 47) to the brightest 1.63$\times 10^{40}$ erg s$^{-1}$ (source No. 42, SN 2014C). There are 11 sources with maximum luminosity $> 10^{39}$ erg s$^{-1}$ in addition to SN 2014C. 

The long term variability of X-ray fluxes of each source is calculated according to the difference between the observed maximum (F$_{max}$) and minimum fluxes (F$_{min}$) during the time span of more than 16 years for this study. We determine the variability parameter S \citep{pri93} as :
\begin{eqnarray}\label{eq:1}
S(F_{max} - F_{min}) = \frac{F_{max}-F_{min}}{\sqrt{\sigma^{2}_{max}+\sigma^{2}_{min}}}
\end{eqnarray}
, where $\sigma_{max}$ and $\sigma_{min}$ are the errors of F$_{max}$ and F$_{min}$ respectively. There are 9 sources with S $>$ 3 that are classified to be variables. However, the above method only compare among the detected sources. For the sources not detected in some observations, such as the cases with zero counts or the net count rate is less than $10^{-6}$, only an upper limit of the flux can be obtained with large uncertainties which makes S be less than 3 but their source significance are larger than 3 in other observations. There are 4 such kind of sources and are also considered as variables. A total of 13 variables, among 48 sources with multiple detections, are denoted as a ``v" in Table~\ref{tab:PointSource_table}. 

We also check if there is any transient in our sample by following the criterion of defining a transient in \cite{lay17}. They define a transient as the source with a maximum to minimum flux ratio $>$ 10 and a flux variability $>$ 3. There are three sources (source No. 38, 45, and 49) that meet these requirements and are denoted in ``T" in Table~\ref{tab:PointSource_table}. 

\subsection{X-ray colors and spectral fitting} \label{subsec:spectra}

Apart from the flux variability, spectral features are also very important to identify the nature of X-ray sources. Due to the short exposure time of all observations, most of the sources are too faint for spectral fit. We chose to calculate their hardness ratios (HR) to get a rough estimation of the spectral properties, which is equivalent to X-ray colors. 

\begin{figure}[ht]
\figurenum{3}
\epsscale{1.2}
\plotone{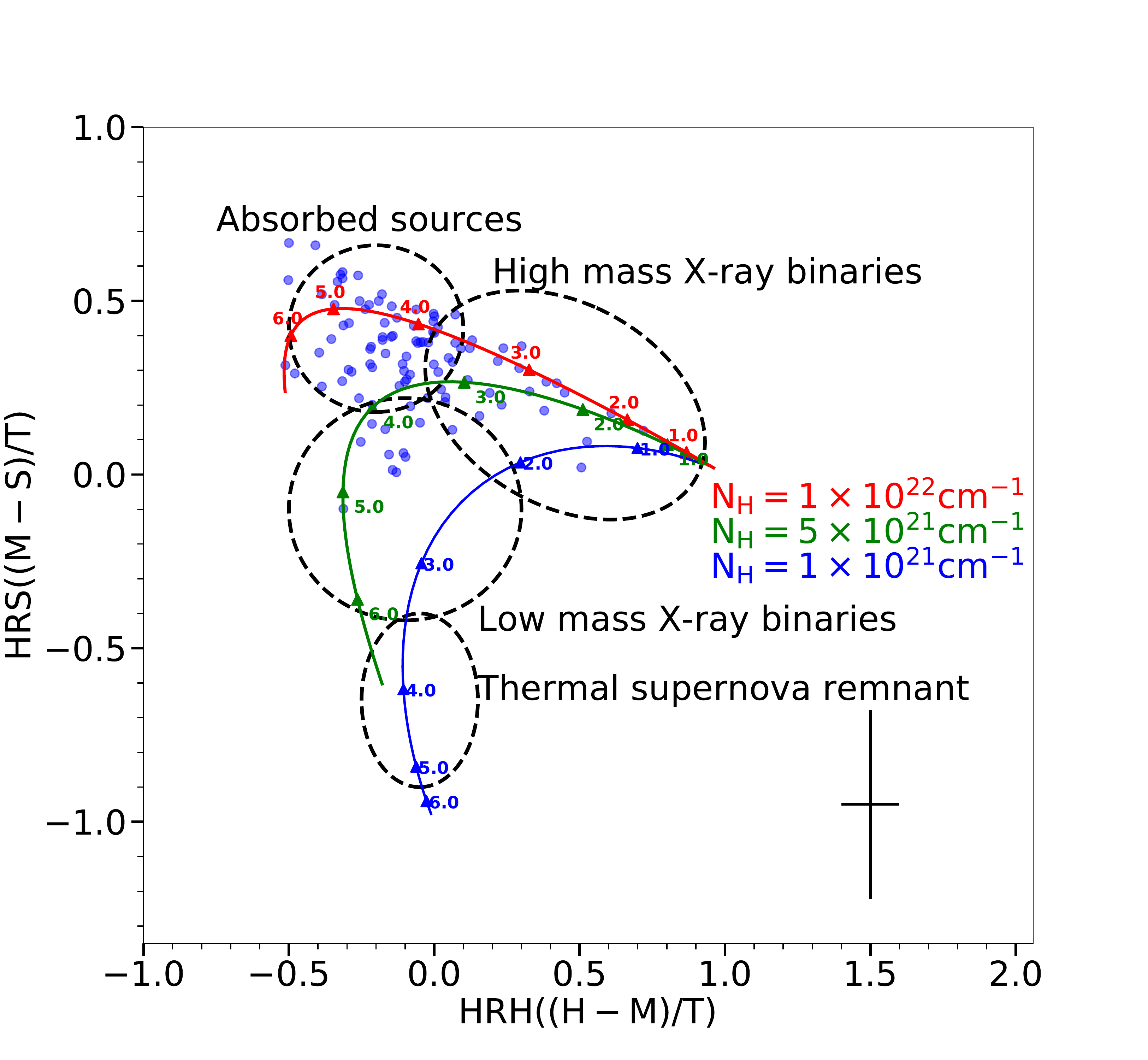}
\caption{X-ray color-color diagram. Soft color HRS = $\left( M - S  \right)/ T $ and hard color HRH = $\left( H - M  \right)/ T$, where S, M, and H are the counts in soft (0.3 -- 1 keV), medium (1 -- 2 keV), and hard (2 -- 8 keV) band. T is the total counts of S, M, and H bands. The light blue dots are all the detections with source significance larger than 4. The four dashed circles are adopted from the classification of X-ray sources in \cite{pre03}. The vertically rising curves show increasing absorption under the assumption of power-law with left to right increasing photon indices. \label{fig3}}
\end{figure}

X-ray colors can be used to classify X-ray sources as discussed in \cite{pre03}. They applied a color-color diagram to discriminate between HMXB, LMXB, and supernova remnant (SNR) candidates. X-ray soft colors are defined, in their paper, as HRS = $\left( M - S  \right)/ T $ and the hard colors as HRH = $\left( H - M  \right)/ T$, where S, M, and H are the counts in soft (0.3 -- 1 keV), medium (1 -- 2 keV), and hard (2 -- 8 keV) band. T is the total counts of S, M, and H bands. In this work, we only consider the sources with source significance larger than 4. Some sources are multiply detected with source significance $>$ 4 and all of them are displayed in Figure~\ref{fig3}. We can classify 4 LMXBs, 4 HMXBs, and 15 absorbed sources, whereas the classification of the remaining 13 are ambiguous because at least one of their detections is located in overlapping regions. Their possible classifications are also listed in Table~\ref{tab:PointSource_table}. 

By comparing the HR variation with the light curve, transitions between hard and soft HR are also found as the flux varies. Particularly, some of the variables (See section~\ref{subsec:flux}) possess bimodal high/soft and low/hard states. When the flux is higher (lower), the HR is softer (harder) which is a typical state transition of LMXBs. Ten sources in our study can be identified as containing the bimodal property. They are all labeled with a letter ``B" in Table~\ref{tab:PointSource_table}. Six of them are classified as either LMXB or HMXB by X-ray color classification method described earlier. This provides additional support that these 6 X-ray sources are very likely to be X-ray binaries.

Because photon counts of each source in individual observation are not sufficient  for a meaningful spectral fit, we therefore investigate the source spectra by stacking all the observations. We first generate separate PHA (Pulse Height Amplitude), RMF, and ARF files for each observation. Then we extract all the seven spectra via the \texttt{specextract} script in \texttt{CIAO} by setting the ``combine" parameter to ``yes". It is the same as running the \texttt{combine$\_$spectra} script in \texttt{CIAO}, which sums up all the source spectra with the exposure-weighted ARFs and ARF and exposure-weighted RMFs of both the source and background spectra. The combined PHA file is grouped with at least 10 counts per bin with \texttt{GRPPHA} tool of \texttt{HEASoft} (version 6.19.)

The spectral fitting of the combined spectra is carried out by \texttt{Sherpa} (version 1) \citep{fre01} with an absorbed power-law model and using its implemented \texttt{Bayesian Low-Count X-ray Spectral} (\texttt{BLoCXS}) to get the lower and upper confidence interval of the fitted parameters. The fit optimization method is set to Nelder-Mead Simplex together with Cash fit statistic \citep{cas79}. We only fit the combined spectra for the 24 sources (except for SN 2014C) that have net counts no less than 15 in their brightest detection, are not variables, and do not possess the bimodal feature. The fitting results of the column density of hydrogen, the photon index of power-law model and the fluxes are listed in Table~\ref{tab:spectral_fitting}. The average best-fit photon index and $N_H$ of the 24 sources are 1.7 and $2.1\times10^{21}$ cm$^{-2}$, respectively (See Table~\ref{tab:spectral_fitting}).

We also investigate the fluxes at the sites of the two historical type II supernovae located within D$_{25}$ of NGC 7331, SN 1959D, and SN 2013bu. The 3$\sigma$ upper limits of both supernovae are 2$\times$10$^{37}$erg s$^{-1}$ by assuming a power-law with Galactic column density and a photon index of 1.7. 

\begin{deluxetable}{c|ccc}
\tablecaption{Spectral Fitting Result of the Bright Sources\label{tab:spectral_fitting}}
\tablenum{3}
\tablehead{
\colhead{}& \multicolumn{3}{c}{Absorbed power-law}
} 
\startdata
Source & $\mathrm{{N}_{H}}$\tablenotemark{a} & $\Gamma$\tablenotemark{b} & Flux (0.3 -- 8 keV)\tablenotemark{c} \\
No.  & ($\mathrm{\times {10}^{22} {cm }^{-2}}$) &  & (erg s$^{-1}$ cm$^{-2}$ ) \\
\cline{1-4}
1 & 0.08 (-----\phantom{.}, -----\phantom{.}) & 1.48 (1.37, 1.76) & 1.57E-14 \\
8 & 0.08 (-----\phantom{.}, -----\phantom{.}) & 0.38 (0.20, 0.80) & 5.43E-15 \\
9 & 0.53 (0.45, 0.97) & 1.43 (1.29, 2.03) & 1.13E-14 \\
11 & 0.71 (0.66, 1.44) & 1.67 (1.59, 2.49) & 7.71E-15 \\
12 & 0.16 (0.13, 0.51) & 1.44 (1.34, 2.23) & 4.25E-15 \\
14 & 0.26 (0.21, 0.51) & 2.30 (2.12, 3.09) & 3.27E-15 \\
17 & 0.40 (0.36, 0.96) & 1.85 (1.74, 2.78) & 6.07E-15 \\
19 & 0.21 (0.17, 0.29) & 1.46 (1.35, 1.64) & 3.62E-14 \\
20 & 0.18 (0.15, 0.25) & 1.93 (1.80, 2.14) & 2.48E-14 \\
22 & 0.08 (-----\phantom{.}, -----\phantom{.}) & 1.69 (1.50, 1.88) & 5.47E-15 \\
23 & 0.08 (-----\phantom{.}, -----\phantom{.}) & 1.85 (1.75, 1.97) & 1.57E-14 \\
24 & 0.08 (-----\phantom{.}, -----\phantom{.}) & 1.58 (1.48, 1.98) & 8.77E-15 \\
27 & 0.08 (-----\phantom{.}, -----\phantom{.}) & 1.63 (1.54, 2.34) & 3.71E-15 \\
28 & 0.11 (0.09, 0.33) & 3.00 (2.84, 4.23) & 2.02E-15 \\
30 & 0.32 (0.32, 0.84) & 2.23 (2.19, 3.45) & 3.15E-15 \\
33 & 0.57 (0.62, 1.69) & 1.64 (1.68, 2.89) & 6.30E-15 \\
34 & 0.10 (0.09, 0.36) & 2.10 (1.99, 3.16) & 2.36E-15 \\
35 & 0.09 (-----\phantom{.}, 0.17) & 1.66 (1.53, 1.94) & 1.37E-14 \\
37 & 0.10 (-----\phantom{.}, 0.30) & 1.04 (0.94, 1.52) & 7.75E-15 \\
39 & 0.08 (-----\phantom{.}, -----\phantom{.}) & 1.49 (1.39, 1.80) & 1.31E-14 \\
41 & 0.08 (-----\phantom{.}, -----\phantom{.}) & 1.09 (0.87, 1.42) & 3.60E-15 \\
43 & 0.46 (0.41, 0.57) & 1.84 (1.72, 2.04) & 3.87E-14 \\
44 & 0.16 (0.11, 0.32) & 1.40 (1.25, 1.81) & 9.25E-15 \\
47 & 0.14 (0.12, 0.44) & 1.51 (1.40, 2.29) & 3.43E-15 
\enddata
\tablenotetext{a}{$\mathrm{{N}_{H}}$: the best fitted value of the column density of hydrogen. For the sources with unconstrained minimum boundary, the $\mathrm{{N}_{H}}$ are fixed to Galactic values.}
\tablenotetext{b}{$\Gamma$: the best fitted value of the photon index of power-law model}
\tablenotetext{c}{ Flux: calculated flux in 0.3 -- 8 keV}
\tablecomments{The fitting in absorbed power-law is based on the combined spectra in 0.3 -- 8 keV energy range. The 68$\%$ confidence ranges of the fitted parameters are listed in the parentheses. This table does not include the spectral fitting result of SN 2014C. The detailed fitting result is stated in Section~\ref{subsec:spectra}.}
\end{deluxetable}

\subsection{X-ray analysis of SN 2014C } \label{subsec:sn2014c}

The first Chandra X-ray observation (ObsID 2198) of the field of SN 2014C was taken before its explosion with an exposure time of 29.5 ks. Assuming an absorbed power-law model with $\Gamma=1.7$, the $3\sigma$ unabsorbed 0.3 -- 8 keV luminosity limit is $8.7\times10^{37}$  erg s$^{-1}$. The first Chandra detection after the explosion was t = 308 d (ObsID 16005) and a series of Chandra and NuSTAR joint observations were performed since t = 397 d. We analyzed five pairs of Chandra and NuSTAR observations in this study from t = 397 d to t = 1029 d (See Table~\ref{tab:SNspec}) with detailed spectral fitting using \texttt{Xspec}. 

The data reduction and spectral analysis of the Chandra data follow the same procedure in Section~\ref{subsec:spectra}. As for the NuSTAR data, the standard pipeline processing is performed to produce the calibrated and cleaned Level 2 event files. The spectra as well as the ARF and RMF files of the specified source and background regions are then created by the \texttt{nuproducts} script. We include the event lists observed by both telescopes of NuSTAR in the spectral analysis. Although NuSTAR covers an energy range of 3 -- 79 keV, the background emission is large and the S/N is low above 30 keV. For Chandra, it is sensitive from 0.3 to 10 keV, but the counts above 8 keV are low. Consequently, we use only the spectra between 3 and 30 keV of NuSTAR and 0.3 to 8 keV of Chandra. All the Chandra and NuSTAR data are grouped to contain at least 30 counts in each bin for spectral fitting.

Owing to the poorer spatial resolution of NuSTAR than Chandra, we can not distinguish the emission of SN 2014C from that of the other nearby X-ray sources located within the Point Spread Function (PSF) of NuSTAR. The contamination of these unresolved sources in the spectra of NuSTAR should be taken into account before the spectral fitting of both instruments. We follow Margutti et al's (2017) method of removing the contamination component from the NuSTAR spectra. First, the spectrum of Chandra is fitted in the region of an annulus centered at the location of SN 2014C with an absorbed power-law model and is extrapolated to the energy range of NuSTAR. The inner 1.5" and outer 1'  radii of the annulus are set according to the PSF of SN 2014C in Chandra and in NuSTAR respectively. Then the best-fit parameters are adopted in a power-law model included in the NuSTAR model fitting for the contamination emission. In other words, we model not only the emission of SN 2014C but also an additional power-law component for the unresolved sources lying within the PSF of NuSTAR.

\begin{figure}
\figurenum{4}
\gridline{
		\includegraphics[trim={0 0 0 0}, clip, angle=270,width=0.48\textwidth]{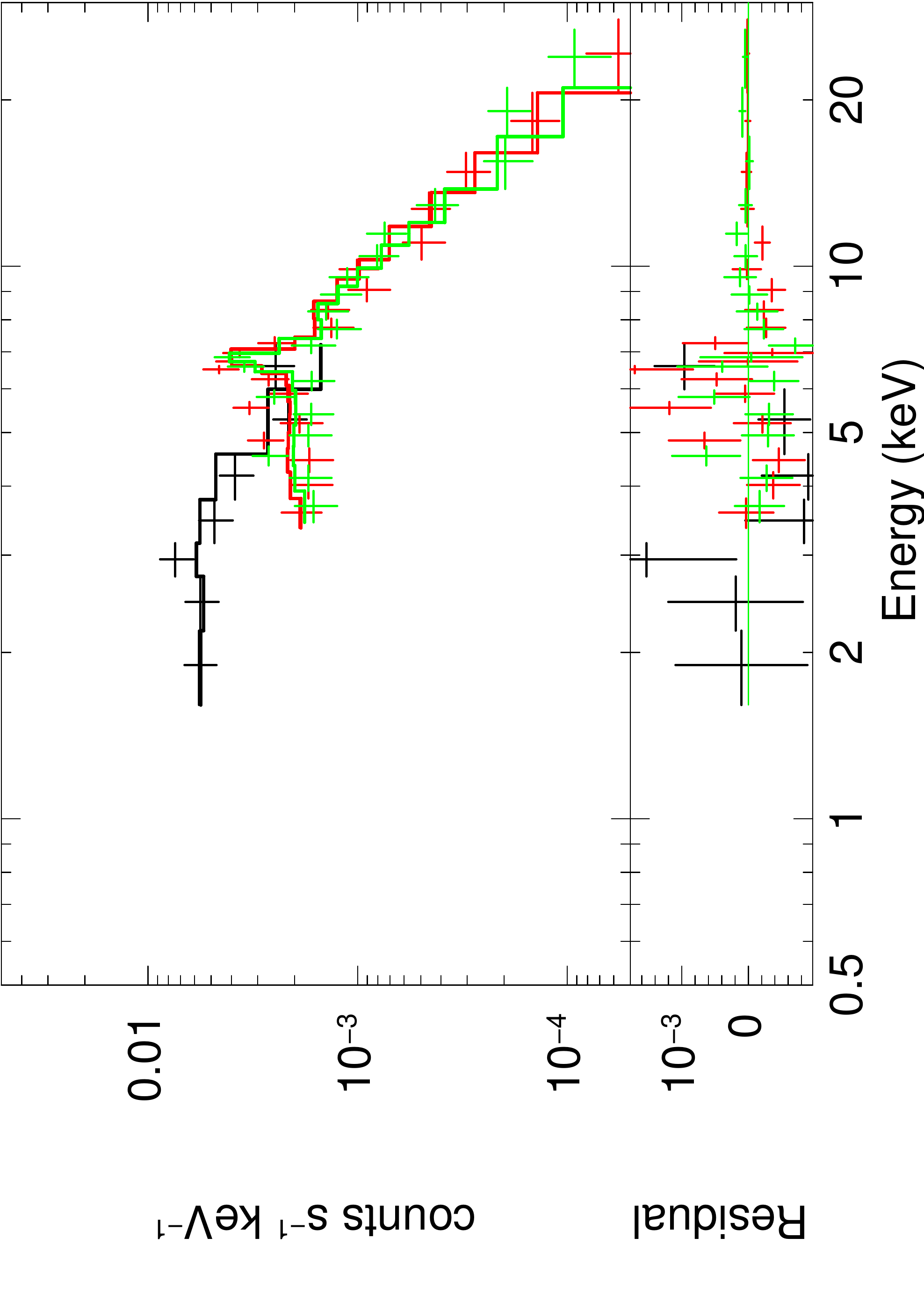}
		}
\gridline{		
	\includegraphics[trim={0 0 0 0}, clip, angle=270,width=0.48\textwidth]{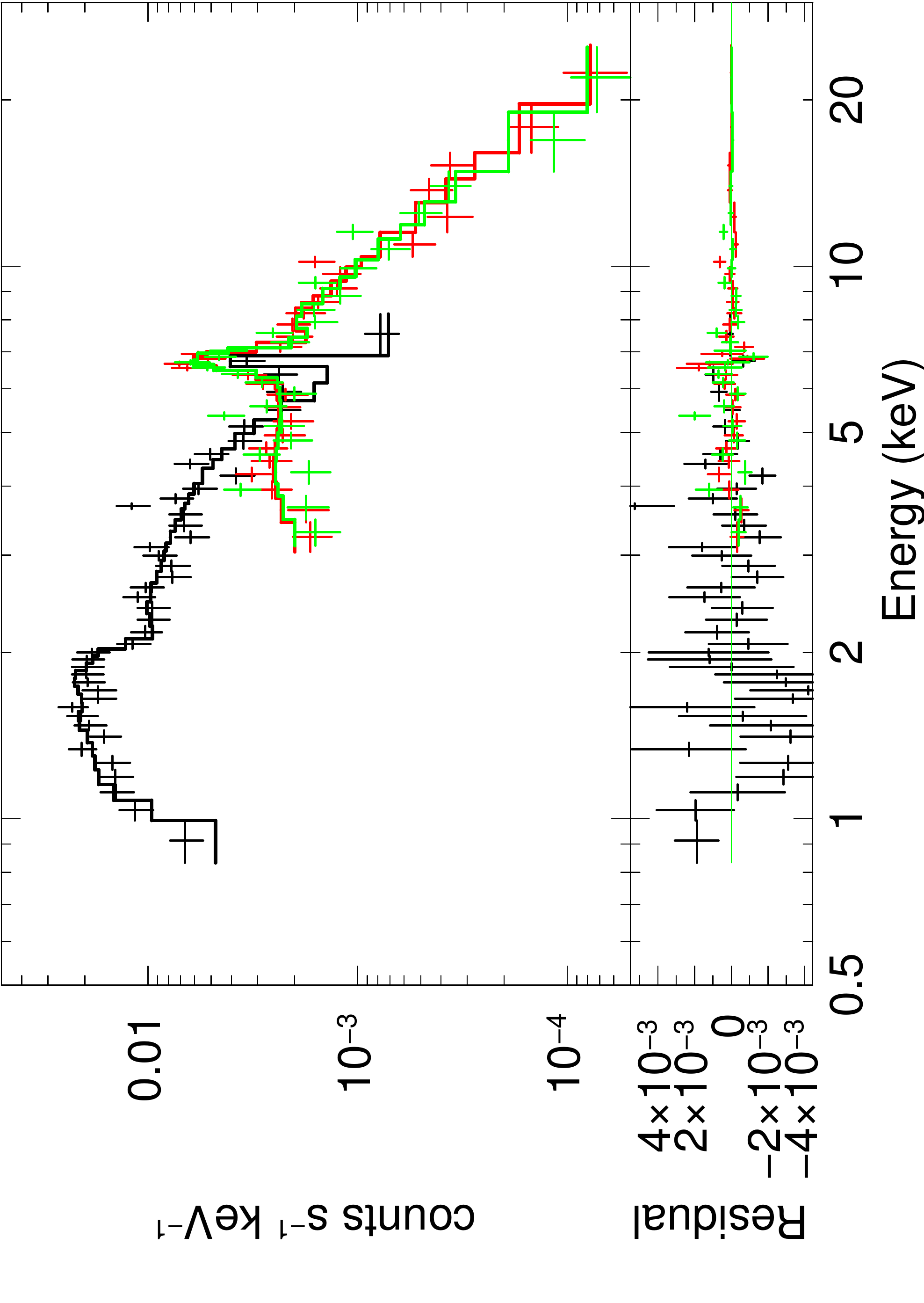}
	}	
\caption{The spectra and their residuals of SN 2014C fitted with an absorbed \texttt{vapec} model at t = 397 d (top panel) and 1029 d (bottom panel) after explosion.\label{fig4}}
\end{figure}

Having taken the contamination into account, we first fit the Chandra-NuSTAR joint spectra with an absorbed power-law model. Since an emission line is apparent in between 6 and 7 keV (See Figure~\ref{fig4}), an extra gaussian component is also added to the model. The best fitted parameters of the five epochs are listed in Table~\ref{tab:SNspec}. The column densities fitted in the absorption component include not only the intrinsic local absorption but also that from Galactic absorption. However, we neglect its contribution since the Galactic $\mathrm{N_{H}} \sim$ 8.62$\mathrm{\times {10}^{20} {cm }^{-2}}$, evaluated by an interactive program \texttt{COLDEN} \citep{sta92, dic90}, is much less than the fitted $\mathrm{N_{H}}$ values in our analysis. The column densities in the power-law model show a clear decreasing trend from the first year after explosion (t = 397 d) to more than three year after explosion (t = 1029 d) due to the expansion of the supernova cloud. On the other hand, the photon-indexes remain to be $\sim$ 1.7 during this period of time. The best fitted central energy of the gaussian profile falls at the range of $\sim$ 6.7 to 6.8 keV with an average equivalent width of 1.35 keV and an average full width half maximum of $\sim$ 0.5 keV.

To obtain the thermal properties of the supernova, we first try to fit the data with an absorbed thermal Bremsstrahlung model added with a gaussian profile. The best fitted plasma temperatures kT are around 20 -- 30 keV but the temperature constraints in the first three data pairs are rather poor for this model (See Table~\ref{tab:SNspec}). As a result, we opt for anther possible thermal spectral model. Due to the inferred high density in the order of $\sim$ 10$^{6}$ cm$^{-3}$) of the H-rich CSM shell \citep{mar17}, the ionization time scale is much shorther than the explosion time discussed in our study. For this reason, we favor an equilibrium ionization model over the non-equilibrium one. An absorbed emission spectrum model, \texttt{vapec}, is then adopted. 

\texttt{vapec} stands for Astrophysical Plasma Emission Code which is an emission spectrum model dealing with the emission from collisionally-ionized hot plasma in ionization equilibrium (CIE) incorporated with the AtomDB atomic database with the variant abundances of C, N, O, Ne, Mg, Al, Si, S, Ca, Ar, Fe, and Ni. The distance adopted in this model is anchored at redshift z = 0.002732 inferred from the Cepheid variables measurement \citep{fre01}. At the explosion site of SN 2014C, \cite{mil15} measured the log(O/H) + 12 = 8.6 $\pm$ 0.1 from their t = 4 day spectrum, which is very close to solar value log(O/H)$_{\sun}$ + 12 = 8.69 \citep{asp05}. Therefore, we set the abundance of all the trace elements to be solar values in the model except for Fe because of a probable iron emission line between 6 -- 7 keV. The abundance table used here is based on \cite{and89}. The best fitted absorption values for the \texttt{vapec} model also follow the same decreasing trend of those in the power-law and the Bremmsstrahlung model as the expansion continues. The plasma temperatures inferred in this model are in the range of $\sim$ 10 to 20 keV. We also found that super-solar abundances of Fe are required for all five epochs in our analysis (See Table~\ref{tab:SNspec}), which will be discussed later in Section~\ref{subsec:sn2014c_2}.

\begin{figure}
\figurenum{5}
\centering
\includegraphics[width=0.5
\textwidth]{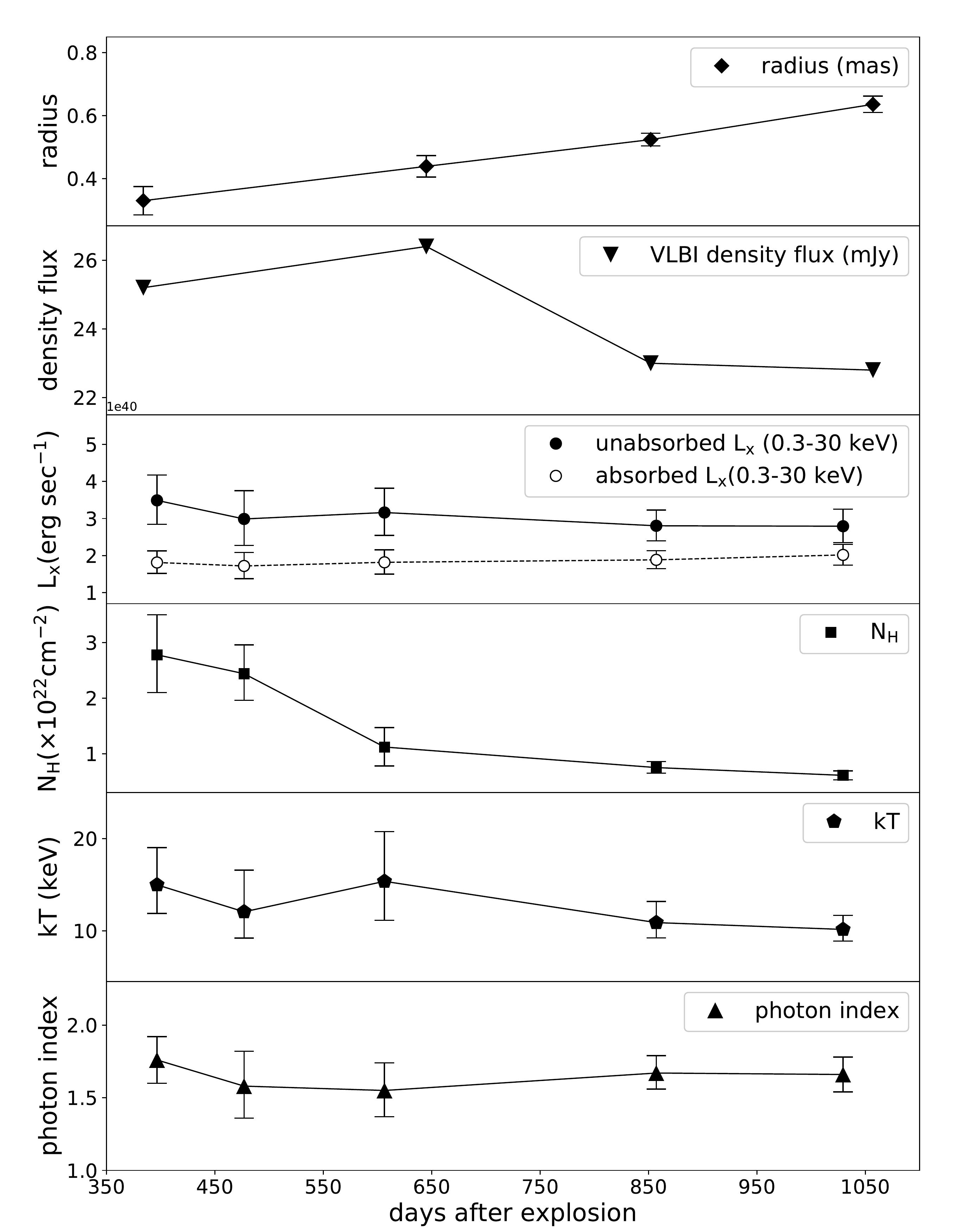}
\caption{Using the Chandra and NuSTAR data, we obtained the X-ray light curve of SN 2014 by fitting a \texttt{tbabs*apec} model in 0.3 -- 30 keV, the variation of column density, the plasma temperature and the photon indexes. The close and open circles represent the unabsorbed and absorbed luminosities respectively. Based on 8.4 GHz VLBI data of SN 2014C, the measured denstiy fluxes and the fitted angular outer radius of a spherical shell model estimated in \cite{bie18} are also plotted for reference. All the errors are at 90$\%$ confidence level.\label{fig5}}
\end{figure}

For the X-ray variability, the 0.3 -- 30 keV luminosity is about $3\times10^{40}$ erg s$^{-1}$ in all epochs within the 90$\%$ uncertainties regardless of the spectral models (see Table~\ref{tab:SNspec} and Figure~\ref{fig5}). However, if we consider the spectra at different epochs, it is obvious that the soft component increases as the expansion continues (See Figure~\ref{fig4}). The absorbed fluxes within 0.3 -- 2 keV increased from 0.94($\pm$0.10)$\times$10$^{-14}$ erg cm$^{-2}$ s$^{-1}$ at t = 397 d to 6.20($\pm$0.27)$\times$10$^{-14}$erg cm$^{-2}$ s$^{-1}$ at t = 1029 d. Fewer soft photons are absorbed when the column densities decrease during the course of two years while the unabsorbed fluxes remain almost constant no matter in soft or hard X-ray.

The detailed analysis of SN 2014C such as its evolution of physical properties and the inferred environment requires a multiwavelength collaborative observation and is beyond the scope of this study.

\begin{deluxetable*}{c|cccccc}
{\tablecaption{Chandra spectral fits of SN 2014C \label{tab:SNspec}}
\tablenum{4}
\tabletypesize{\footnotesize}
\tablewidth{0pt}
\startdata
\tablehead{
\colhead{}&\multicolumn{6}{c}{tbabs*(power-law+gaussian)} \\
\cline{1-7}
Chandra/NuSTAR ObsID (day)\tablenotemark{a} & ${\mathrm{\chi}^{2}_{\nu}}/dof$\tablenotemark{a}  & $\mathrm{{N}_{H}}$\tablenotemark{b}  &  PI\tablenotemark{c} &  LineE\tablenotemark{d} & Flux\tablenotemark{e} & L$_{X}$\tablenotemark{e}  \\
\cline{1-7}
17569/80001085002 (\phantom{.}397\phantom{.}) & 0.88/39 & 2.92($_{-0.84}^{+0.89}$) & 1.76($_{-0.16}^{+0.16}$) & 6.75($_{-0.09}^{+0.09}$) & 1.43($_{-0.29}^{-0.31}$) & 3.70($_{-0.75}^{+0.80}$) \\
17570/40102014001 (\phantom{.}477\phantom{.}) & 0.95/27 & 2.42($_{-0.63}^{+0.68}$) & 1.58($_{-0.22}^{+0.24}$) & 6.82($_{-0.13}^{+0.18}$) & 1.69($_{-0.35}^{-0.37}$) & 4.36($_{-0.91}^{+0.96}$) \\
17571/40102014003 (\phantom{.}606\phantom{.}) & 1.48/43 & 1.25($_{-0.44}^{+0.46}$) & 1.55($_{-0.18}^{+0.19}$) & 6.76($_{-0.08}^{+0.08}$) & 1.76($_{-0.30}^{-0.31}$) & 4.56($_{-0.77}^{+0.80}$) \\
18340/40202013002 (\phantom{.}857\phantom{.}) & 1.61/88 & 0.86($_{-0.14}^{+0.14}$) & 1.67($_{-0.11}^{+0.12}$) & 6.69($_{-0.10}^{+0.10}$) & 1.79($_{-0.23}^{-0.23}$) & 4.64($_{-0.59}^{+0.61}$) \\
18341/40202013004 (1029) & 1.36/93 & 0.69($_{-0.11}^{+0.12}$) & 1.66($_{-0.12}^{+0.12}$) & 6.73($_{-0.05}^{+0.05}$) & 1.79($_{-0.24}^{-0.25}$) & 4.62($_{-0.63}^{+0.65}$) \\
\cline{1-7}
\colhead{}& \multicolumn{6}{c}{tbabs*(Bremsstrahlung+gaussian)} 
}
Chandra/NuSTAR ObsID (day)\tablenotemark{a} & ${\mathrm{\chi}^{2}_{\nu}}/dof$\tablenotemark{a}  & $\mathrm{{N}_{H}}$\tablenotemark{b}  &  kT(keV) & LineE\tablenotemark{d} & Flux\tablenotemark{e}  & L$_{X}$\tablenotemark{e}  \\
\cline{1-7}
17569/80001085002 (\phantom{.}397\phantom{.}) & 0.91/39 & 2.14($_{-0.72}^{+0.76}$) & 24.04($_{-\phantom{0}6.72}^{+12.71}$) & 6.75($_{-0.09}^{+0.09}$) & 1.42($_{-0.24}^{+0.26}$) & 3.67($_{-0.63}^{+0.67}$) \\
17570/40102014001 (\phantom{.}477\phantom{.}) & 0.86/27 & 2.03($_{-0.50}^{+0.55}$) & 23.87($_{-\phantom{0}9.22}^{+25.24}$) & 6.81($_{-0.12}^{+0.17}$) & 1.51($_{-0.31}^{+0.33}$) & 3.91($_{-0.79}^{+0.84}$) \\
17571/40102014003 (\phantom{.}606\phantom{.}) & 1.43/43 & 0.97($_{-0.37}^{+0.38}$) & 34.97($_{-14.20}^{+42.93}$) & 6.76($_{-0.08}^{+0.08}$) & 1.69($_{-0.27}^{+0.29}$) & 4.37($_{-0.71}^{+0.75}$) \\
18340/40202013002 (\phantom{.}857\phantom{.}) & 1.24/88 & 0.67($_{-0.10}^{+0.11}$) & 16.00($_{-\phantom{0}3.74}^{+\phantom{0}5.03}$) & 6.69($_{-0.10}^{+0.10}$) & 1.34($_{-0.18}^{+0.18}$) & 3.46($_{-0.45}^{+0.47}$) \\
18341/40202013004 (1029) & 1.11/93 & 0.53($_{-0.08}^{+0.09}$) & 15.85($_{-\phantom{0}4.09}^{+\phantom{0}5.84}$) & 6.73($_{-0.05}^{+0.05}$) & 1.31($_{-0.18}^{+0.19}$) & 3.38($_{-0.48}^{+0.50}$) \\
\cline{1-7}
\colhead{}&\multicolumn{6}{c}{tbabs*vapec} \\
\cline{1-7}
Chandra/NuSTAR ObsID (day)\tablenotemark{a} & ${\mathrm{\chi}^{2}_{\nu}}/dof$\tablenotemark{a}  & $\mathrm{{N}_{H}}$\tablenotemark{b}  &  kT(keV) & Fe abundance\tablenotemark{f} & Flux\tablenotemark{e}  & L$_{X}$\tablenotemark{e}  \\
\cline{1-7}
17569/80001085002 (\phantom{.}397\phantom{.}) & 1.21/42 & 2.78($_{-0.68}^{+0.72}$) & 14.99($_{-3.10}^{+4.03}$) & 2.54($_{-0.98}^{+1.42}$) & 1.35($_{-0.25}^{+0.27}$) & 3.49($_{-0.64}^{+0.69}$) \\
17570/40102014001 (\phantom{.}477\phantom{.}) & 0.99/30 & 2.44($_{-0.48}^{+0.52}$) & 12.06($_{-2.85}^{+4.53}$) & 2.83($_{-1.16}^{+2.17}$)  & 1.16($_{-0.28}^{+0.30}$) & 2.99($_{-0.72}^{+0.77}$) \\
17571/40102014003 (\phantom{.}606\phantom{.}) & 1.48/46 & 1.12($_{-0.34}^{+0.35}$) & 15.37($_{-4.23}^{+5.39}$) & 4.73($_{-1.92}^{+3.09}$)  & 1.22($_{-0.24}^{+0.25}$) & 3.16($_{-0.62}^{+0.66}$) \\
18340/40202013002 (\phantom{.}857\phantom{.}) & 1.23/91 & 0.75($_{-0.10}^{+0.11}$) & 10.89($_{-1.65}^{+2.30}$) & 1.79($_{-0.54}^{+0.75}$)  & 1.08($_{-0.16}^{+0.16}$) & 2.80($_{-0.41}^{+0.42}$) \\
18341/40202013004 (1029) & 1.13/96 & 0.61($_{-0.08}^{+0.08}$) & 10.15($_{-1.27}^{+1.52}$) & 3.70($_{-0.84}^{+1.04}$)  & 1.08($_{-0.17}^{+0.18}$) & 2.79($_{-0.44}^{+0.46}$) \\
\cline{1-7}
\enddata
}
{
\tablenotetext{a}{day: number of days after explosion based on Chandra observation date and the gap between Chandra and NuSTAR observing time is within a week; ${\mathrm{\chi}^{2}_{\nu}}$: reduced chi-square; dof: degrees of freedom}
\tablenotetext{b}{$\mathrm{{N}_{H}}$: column density of hydrogen in units of $\mathrm{\times {10}^{22} {cm }^{-2}}$}
\tablenotetext{c}{PI: power-law index.}
\tablenotetext{d}{LineE: central energy of the Gaussain profile in keV.}
\tablenotetext{e}{Flux: 0.3 -- 30 keV unabsorbed flux in $\times$10$^{-12}$ erg cm$^{-2}$ s$^{-1}$; L$_{X}$: X-ray luminosity of 0.3 -- 30 keV in $\times$10$^{40}$ erg s$^{-1}$. }
\tablenotetext{f}{Fe abundance: in the unit of solar abundance.}
\tablecomments{
The fitting is based on the spectra of SN 2014C in the energy range of 0.3 -- 8 keV for Chandra and 3 -- 30 keV for NuSTAR observations. The 90$\%$ confidence errors are listed in the parentheses.}
}
\end{deluxetable*}

\subsection{X-ray luminosity function} \label{subsec:xlf}

X-ray luminosity functions (XLFs) of point sources can reflect the stellar population properties and the evolution of their host galaxies \citep{fab06_2, kon03_2}. However, the nature of variability of X-ray compact sources may also affect their XLFs. Some studies find XLFs are quite stable \citep{zez04, gri05}, but \cite{bin17} claim that XLFs differ from observation to observation. Therefore, our study is of great interest, not only to reveal the X-ray population properties of NGC 7331 but also to examine whether its XLFs are steady based on the seven Chandra observations spanning over 15 years. 

The luminosities of X-ray point sources are estimated by applying an absorbed power-law with $\Gamma$= 1.7 and the Galactic N$_{H}$ in the 0.3 -- 8 keV  band. Only the sources with detection significance larger than 3 are included in the XLF analysis. The faintest sources suffering incompleteness are removed from the list. The incompleteness correction is carried out by using MARX \citep{dav12} to simulate a ray-trace reprojection onto the Chandra detector plane. We make use of the spectra of the available detected sources in the previous steps in Section~\ref{subsec:flux} for the MARX simulation. In order to take the spatial diffuse background and the PSF degradation of off-AXIS into account, we estimate the detection probability in different off-axis position or background environment. Sources with lower than 99$\%$ detection probability are excluded in the XLFs fitting. We also exclude the luminous SN 2014 and the sources with equivalent luminosity in the following XLF fitting. The luminosities of the sources included for XLF fitting are in the range from 1.18$\times$10$^{38}$ to 2.36$\times$10$^{39}$ erg s$^{-1}$. The cumulative XLFs of each observation is plotted in colored closed circles in Figure~\ref{fig6}.

We follow the majority of earlier literature to fit the XLFs with a simple power-law (PL) model as well as a power-law model with an exponential cut-off (PLC). They are in the forms of: 
\begin{align}
N(>L) &= K_{1}L ^{ -\alpha_{1}} \label{eq:2}
\intertext{and}
N(>L) &= K_{2}L^{ -\alpha_{2}}exp(\frac{-L}{L_{c}})  \label{eq:3}
\end{align}
respectively. Here, N is the accumulated number of X-ray sources larger than the luminosity L, both $\mathrm{K}_{1}$ and $\mathrm{K}_{2}$ are the normalizations and $L_{c}$ is the cut-off luminosity. $\alpha_{1}$ and $\alpha_{2}$ are the power indexes which also are the slopes of the log-scaled cumulative XLFs.

Instead of using the traditionally adopted frequentist inference, we opt to Bayesian method to model the XLFs. Although both the classical frequentist and Bayesian inference are designed to find the highest data likelihood of each parameter, the former one cannot tell us the probability density function (PDF) for model parameters but the later can do. The difference between these two statistical analysis approaches is often negligible in large data sets but obvious when the number statistic is low, such as the XLF fitting in this study. We use \texttt{PyMC3} \citep{sal16} package, a probabilistic programming for Bayesian statistical modeling written in Python, for the XLF fitting. \texttt{PyMC3} implements advanced Markov chain Monte Carlo (MCMC) sampling algorithms with user-defined probabilistic models for Bayesian inference. A self-tuning variant of Hamiltonian Monte Carlo (HMC) \citep{dua87} called ``The No-U-Turn Sampler (NUTS)" \citep{hof14} is adopted in sampling process in the fitting. 

First, we fit the individual XLFs of each observation with PL and PLC model. Both the means of the posteriors of power index ($\alpha_{1}$ in Equation~\ref{eq:2} and $\alpha_{2}$ in Equation~\ref{eq:3}) of individual datasets are rather scattered so are their posterior predictive distributions as shown in Figure~\ref{fig6} (in violet). The posterior means of the power index range from 1.06 (HPD = 0.97 -- 1.15) to 1.42 (HPD = 1.28 -- 1.53) for PL model and from 0.64 (HPD = 0.59 -- 0.69) to 1.19 (HPD = 1.12 -- 1.25) for PLC model (See Table~\ref{tab:xlf_fitting}). The scattering is very likely due to the varying luminous sources among the different observations. A fraction of X-ray sources in NGC 7331 are variables and transients as we stated in Section~\ref{subsec:flux}. This is the primary reason causing the variability of XLFs and these variables would cause more obvious deviation for the cases with sparse detected sources. Moreover, the best cut-off luminosity is not constrained for the last 6 datasets, showing that PLC is not a proper XLF model for individual dataset.
 
If we wish to get a more ``representative" XLF of a galaxy, another way is to model the XLF with all the datasets being pooled together. This is equivalent to combining all the data from different observations. With multiple observations, the impact of the high luminosity tail in each observation would be less in a bigger data pool. For the pooled XLF fitting, the posterior mean of the simple power-law slope ($\alpha_{1}$) is 1.18 with the 99$\%$ highest posterior density (HPD) from 1.12 to 1.24, which is in the range of those based on individual observations. We can also fit the pooled data with a cut-off power-law with a slope ($\alpha_{2}$) of 0.62 (HPD = 0.49 -- 0.75) and a mean cut-off luminosity of  1.05 (HPD = 0.89 -- 1.31)$\times$10$^{39}$ erg s$^{-1}$, that is very close to the fit of ObsID 02198. This is because ObsID02198 has the largest number of detected sources due to its long exposure time. Moreover, the PLC model of the remaining datasets are not well constrained owing to the limited number of sources. This implies that observations with more detected sources would dominate the XLF fitting with pooled method because we treat all sources as a whole without considering the differences among observations.

\begin{figure}
\figurenum{6}
\includegraphics[width=1.0\columnwidth]{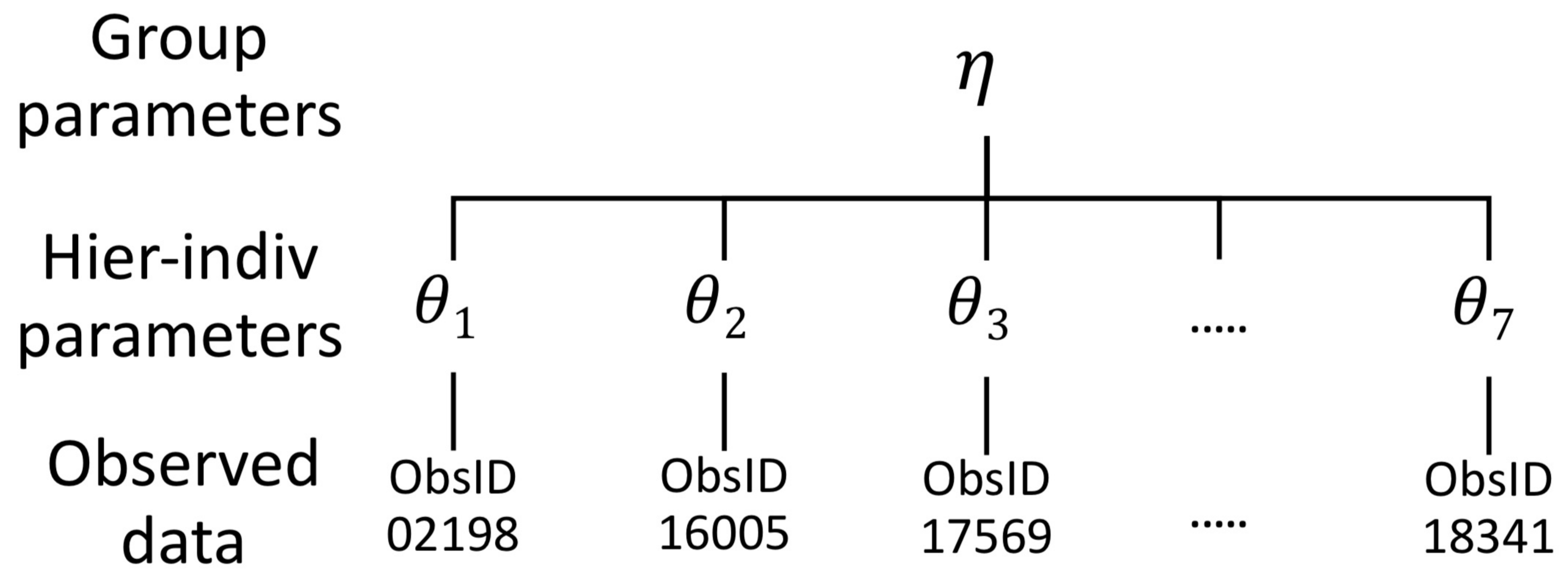}
\caption{Hierarchical model structure.} \label{fig:hier_structure}
\end{figure}

\begin{figure*}
\figurenum{7}
\gridline{\fig{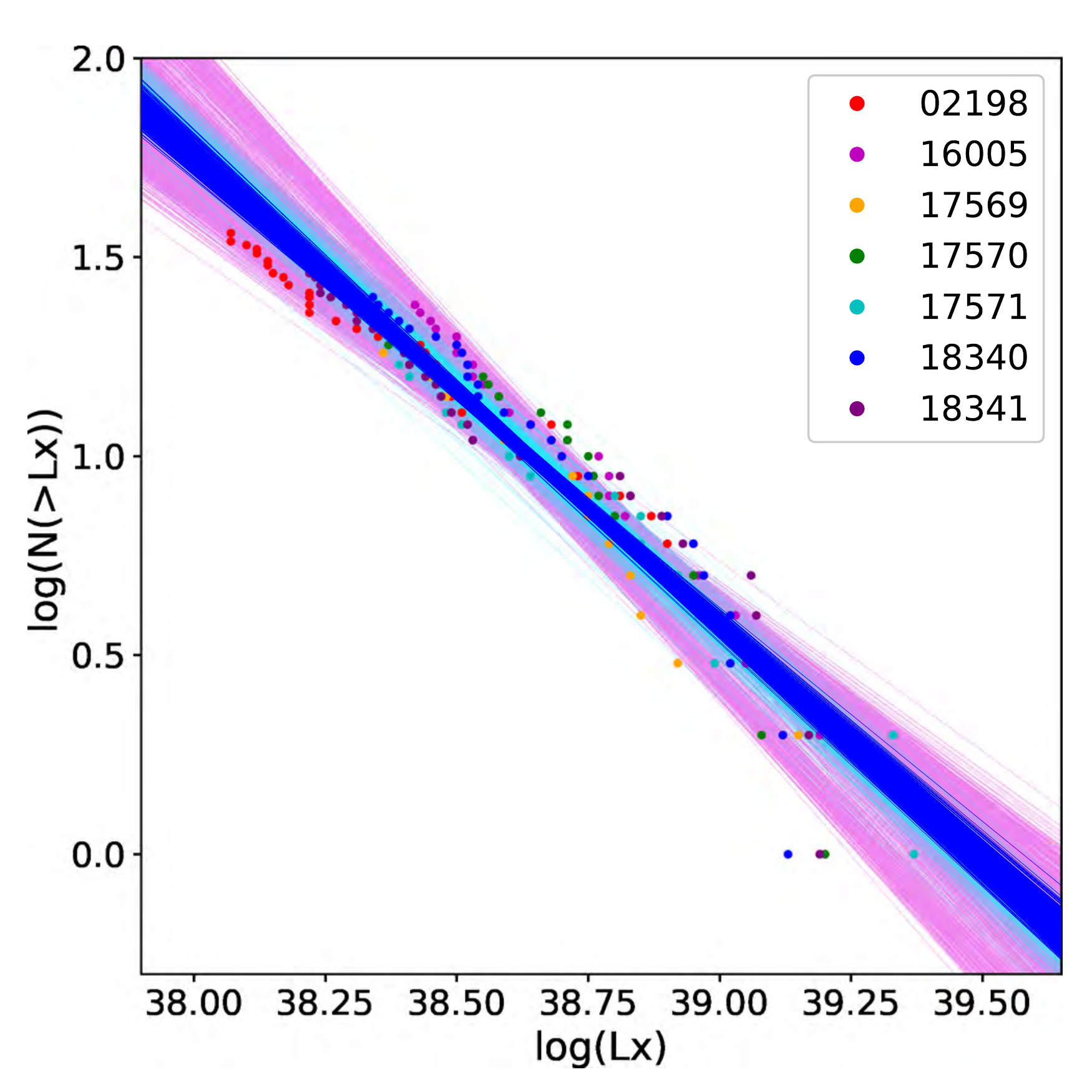}{0.45\textwidth}{(a)\label{fig6a}}
				\fig{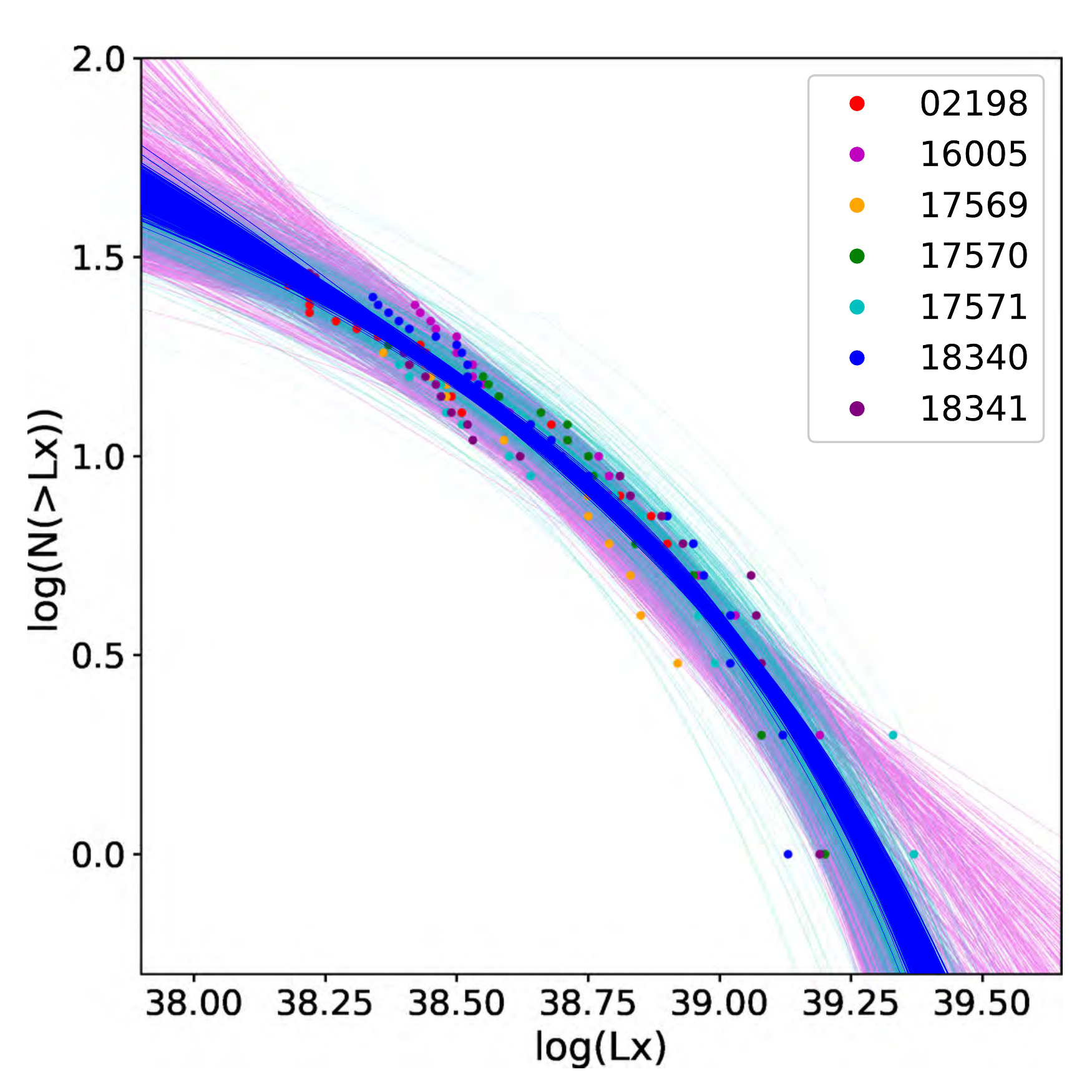}{0.45\textwidth}{(b)\label{fig6b}}		
          }
\caption{Incompleteness corrected cumulative XLFs and their posterior predictive distributions of the power-law (left) and the power-law with an exponential cut-off (right) fitting. The color dots are the 0.3 -- 8 keV luminosities for the individual observation with the measurements of SN 2014C excluded. The colored thin lines are the posterior predictive distribution of the fitting of the individual observation (violet), hierarchical model (cyan) and pooled data (blue) respectively.} \label{fig6}
\end{figure*}

A better solution to the problems in individual and pooling method is using a hierarchical (also called ``partial pooling") algorithm. Hierarchical method assumes individual observations sharing similarity rather than being totally different. This can be done by assuming the ``hier-indiv" parameters (''$\theta_{i}$" in Figure~\ref{fig:hier_structure}) of each XLF  that are from a common \texttt{group} distribution (``$\eta$" in Figure~\ref{fig:hier_structure}), which in return influences the distribution of the parameters of each dataset (``\texttt{hier-indiv}" parameters). The \texttt{group} and the \texttt{hier-indiv} means of $\alpha_{1}$ of the simple power-law are all 1.21 (HPD = 1.15 -- 1.26), which is close to the pooled result. For the PLC fitting with the hierarchical model, both the mean $\alpha_{2}$ of the \texttt{group} and the \texttt{hier-indivs} have the same value of 0.48 (HPD = 0.36 -- 0.60). The cut-off luminsity of \texttt{group} is 0.79 (HPD = 0.58 -- 1.09)$\times$10$^{39}$ erg s$^{-1}$ and those of the \texttt{hier-indivs} are in the range of $\sim$ 0.66$\times$10$^{39}$(HPD = 0.56 -- 0.79) to 1.89$\times$10$^{39}$ erg s$^{-1}$ (HPD = 1.66 -- 2.12) (See Table~\ref{tab:xlf_fitting}). The results of hierarchical fitting reveal that variability of XLF at different epochs is not prominent statistically anymore as it is in individual fitting even though we do not treat all the observations as independent ones. The results of hierarchical method provides an accountable way of obtaining a representative of XLFs without ignoring the differences among different datasets. 

\begin{deluxetable}{c|c|c|cc}
\tablecaption{X-ray Luminosity Function Fitting Result\label{tab:xlf_fitting}}
\tablenum{5}
\tabletypesize{\footnotesize}
\tablewidth{0pt}
\tablehead{
\colhead{}& \colhead{}& \colhead{power-law}& \multicolumn{2}{c}{power-law with a cut-off}
} 
\startdata
Method & ObsID &  $\mathrm{{\alpha}_{1}}$\tablenotemark{a} & $\mathrm{{\alpha}_{2}}$\tablenotemark{a} & $\mathrm{{L}_{c}}$(10$^{39}$ erg s$^{-1}$)\tablenotemark{a} \\
\cline{1-5}
Individual & 02198 & 1.06	($_{-0.09}^{+0.09}$)	&  0.64 ($_{-0.05}^{+0.05}$)	 &  1.05	($_{-0.05}^{+0.10}$) \\
			&  16005 &  1.38 ($_{-0.10}^{+0.13}$) &   1.19 ($_{-0.07}^{+0.06}$)	 &  ----------- \\
			&  17569 &  1.40 ($_{-0.13}^{+0.13}$) &   0.87 ($_{-0.08}^{+0.07}$)	 &  -----------  \\
			&  17570 &  1.42 ($_{-0.14}^{+0.11}$) &   0.99 ($_{-0.15}^{+0.15}$)	 &  -----------  \\
			&  17571 &  1.09	 ($_{-0.14}^{+0.14}$) &   0.81 ($_{-0.08}^{+0.09}$)	 &  -----------  \\
			&  18340 &  1.32 ($_{-0.14}^{+0.16}$) &   0.86 ($_{-0.10}^{+0.09}$)	 &  -----------  \\
			&  18341 &  1.08 ($_{-0.12}^{+0.12}$) &   0.79 ($_{-0.04}^{+0.05}$)	 &  -----------  \\
\cline{1-5}
Pooled & all  & 1.18 ($_{-0.06}^{+0.06}$)	 & 0.62	 ($_{-0.13}^{+0.13}$)	 & 1.05 ($_{-0.19}^{+0.26}$) \\
\cline{1-5}
Hierarchical & group & 1.21	($_{-0.06}^{+0.01}$)	 &  0.48 	($_{-0.12}^{+0.12}$)	 & 0.79	($_{-0.21}^{+0.30}$) \\
	 & 02198	 & 1.21	($_{-0.05}^{+0.05}$)	 &  0.48 ($_{-0.13}^{+0.12}$)	 & 0.72	($_{-0.12}^{+0.19}$) \\
	 & 16005	 & 1.21	($_{-0.06}^{+0.05}$)	 &  0.48 ($_{-0.13}^{+0.12}$)	 & 0.72	($_{-0.11}	^{+0.15}$) \\
	 & 17569	 & 1.21	($_{-0.05}^{+0.05}$)	 &  0.48 ($_{-0.12}^{+0.12}$)	 & 0.72	($_{-0.12}^{+0.17}$) \\
	 & 17570	 & 1.21	($_{-0.06}^{+0.05}$)	 &  0.48 ($_{-0.12}^{+0.12}$)	 & 0.68	($_{-0.11}^{+0.14}$) \\
	 & 17571	 & 1.21	($_{-0.06}^{+0.05}$)	 &  0.48 ($_{-0.12}^{+0.12}$)	 & 1.35 ($_{-0.28}^{+0.39}$) \\
	 & 18340	 & 1.21	($_{-0.06}^{+0.05}$)	 &  0.48 ($_{-0.13}^{+0.12}$)	 & 0.66	($_{-0.10}^{+0.13}$) \\
	 & 18341	 & 1.21	($_{-0.06}^{+0.05}$)	 &  0.48 ($_{-0.12}^{+0.12}$)	 & 1.89	($_{-0.17}^{+0.23}$) \\
\enddata
\tablenotetext{a}{$\mathrm{\alpha_{1}}$: the power index of power-law model; $\mathrm{\alpha_{2}}$: the power index of power-law with an exponential model; $\mathrm{L_{c}}$: the cut-off luminosity of power-law with an exponential cut-off model. The $\mathrm{L_{c}}$s of the last 6 datasets derived by using individual method are not well constrained.}
\tablecomments{The luminosities are calculated based on an absorbed power-law with Galactic N$_{H}$ in 0.3 -- 8 keV. The 2.5$\%$ and 97.5$\%$ highest posterior density (HPD) of each parameter are listed in the parentheses.}
\end{deluxetable}

The posterior predictive distributions (PPDs) of the power-law and the power-law with a cut-off model fitting in the individual, pooled and hierarchical method are plotted in violet, cyan, and blue thin lines respectively in Figure~\ref{fig6}. Since the PPD of individual algorithm differs from observation to observation, we can see from Figure~\ref{fig6} that the ensemble of them (violet curves) is more divergent than the PPD of pooling method (blue curves). The result does not differ much between pooling and hierarchical method and their PPD coverage coincides with each other as shown in Figure~\ref{fig6}. The priors of the individual, pooled and the \texttt{group} parameter in hierarchical method are assumed to be uniform distribution with boundaries of 1.0 and 2.0 for the power-index and 10$^{37}$ and 10$^{41}$erg s$^{-1}$ for the cut-off luminosity for both PL and PLC model. The priors for the \texttt{group} parameter $\eta$ are set to be in an uninformative uniform distribution and those for the \texttt{hier-indiv} ($\theta_{i}$) are assumed to be a Guassian distribution with the means obtained from a random samples of the \texttt{group} parameter. All the likelihood sampling are set to be in a ``Student's t" distribution.

For comparing the resulting fit of the Bayesian models of XLFs, we utilize the Watanabe-Akaike Information Criterion (WAIC) \citep{wat10, gel14}. WAIC estimates the out-of-sample prediction by computing log pointwise posterior predictive density (LPPD). LPPD sums over the log of the mean of the likelihoods  evaluated at the posterior simulations of the parameter values of each sample. It tells how well each estimate of the parameters from the posterior distribution does in predicting the data. WAIC also includes a penalty term for overfitting of the model. This term estimates the effective number of free parameters in the model by summing up the variance associated with the log posterior predictive density. The lower the WAIC is, the better the model fit is. According to our calculation of WAIC, the hierarchical model can describe the XLF of NGC7331 better than the pooled ones regardless of the functional form. The WAIC results also suggest that power-law with an exponential cut-off model fitted better than a simple power-law due to the lower counts in the high luminosity tail. We can also compare different models with the Akaike weight \citep{wag04}, which can be interpreted as the probability of one model being in favor among all compared models. The Akaike weights are 0.89, 0.08, 0.03, and 0 for the hierarchical PLC, pooled PLC, pooled power-law and hierarchical power-law model respectively. It suggests that a power-law with an exponential cut-off using the hierarchical method can best represent the XLF of NGC 7331.

\section{Optical observations} \label{sec:optical}

In order to search for optical counterparts of the X-ray sources found in NGC 7331, we select five Hubble Space Telescope observations from 1994 to 2015 that contain its D$_{25}$ isophote region. There are three HST Wide Field Planetary Camera-2 (WFPC2) observations (ObsID 07450, 11128, and 11966) and two Wide Field Camera 3 (WFC3) observations (ObsID 14202 and 14668). The observation date, filters, and exposure time of all the HST observations are listed in Table~\ref{tab:HSTLog}. 

\begin{deluxetable}{cccl}[h]
\tablecaption{HST Observation Log \label{tab:HSTLog}}
\tablecolumns{4}
\tablewidth{5pt}
\tablehead{
\colhead{Camera} & 
\colhead{ObsID} & 
\colhead{Date} &
\colhead{Filter} \\
[-0.7cm]\\
\colhead{} & 
\colhead{} & 
\colhead{(yyyy-mm-dd)} &
\colhead{Exposure} 
}
\startdata
WFPC2 & 07450 & 1997-08-13 & F450W (0.38 ks) \\
& & & F814W (0.17 ks)  \\
WFPC2& 11128 & 2007-09-16 & F336W (3.60 ks) \\
& & & F555W (0.40 ks) \\
WFPC2& 11966 & 2009-01-01 & F658\phantom{.}N (1.80 ks) \\
\hline
WFC3 & 14202 & 2015-08-22 & F275W (1.94 ks) \\
 & & & F438W (1.38 ks) \\
& & & F814W (1.35 ks) \\
WFC3 & 14668 & 2016-10-12 & F336W (0.39 ks) \\
& & & F555W (0.71 ks) \\
\enddata
\end{deluxetable}

\subsection{Photometry} \label{subsec:photometry}

The source detection and photometry of HST images are performed by adopting \texttt{dolphot} package, which is a modified version of \texttt{HSTphot} \citep{dol00} to offer a combined photometry list. \texttt{dolphot} is capable of running the photometry with multiple-images and calculating the offset and rotation of the images to align them with each other.

The preparation steps for \texttt{dolphot} photometry include masking out all pixels flagged as bad in the data quality images and creating a sky image to provide the sky maps according to \texttt{dolphot} parameters. Then we perform \texttt{dolphot} on the single-chip format images to create an output photometry list containing the X, Y positions, magnitudes, and signal-to-noise ratios (S/N) of the detected stars. Sources less than 3$\sigma$ detection limit, in hot pixels or are extended sources are identified and rejected by \texttt{dolphot}.

\subsection{Astrometry} \label{subsec:astrometry}

For searching the optical counterpart candidates, we need to improve the relative astrometry by aligning the Chandra X-ray images and the HST ones onto the same reference frame. The reference frame that we adopted is the United States Naval Observatory catalog (USNO-B1.0). We check a wide field image obtained from the ESO-DSS image server and choose reference stars that have corresponding bright sources in the UNSO-B1.0 catalog.

For the optical astrometry, we look up the reference stars appearing in both HST images and in the USNO-B1.0 catalog. By applying the \texttt{IRAF} task \texttt{ccmap} on the HST images, we transform all the HST image coordinates onto the same coordinate frame based on the USNO-B1.0 catalog. Nine optical images (see Table~\ref{tab:HSTLog}) taken with WFPC2 and WFC3 are checked in this study. The registration errors of these 9 HST images range from 0\farcs13 to 0\farcs25 in RA and 0\farcs19 to 0\farcs28 in DEC.

To align the Chandra combined image onto the reference frame, we cross match the X-ray source positions obtained by \texttt{wavdetect} with stars in the USNO-B1.0 catalog. Due to the severe absorption in the central part of the galaxy, there are only four or five matching pairs can be found in each Chandra image. The resultant registration errors are about 0\farcs30 in RA and 0\farcs34 in declination. The X-ray source positions are also shifted accordingly by -0\farcs48 in RA and -0\farcs96 in DEC. As for the position statistical uncertainty of the X-ray sources, it can be obtained from the \texttt{wavdetect} result. It ranges from 0\farcs1 to 0\farcs68 in RA and 0\farcs1 to 0\farcs74. in declination. 

The 1$\sigma$ error position uncertainties of X-ray sources are determined according to the root of the quadratic sum of errors obtained from \texttt{wavdetect}, the X-ray registration errors, and the optical astrometry uncertainties. We consider only the HST sources lying within the 2.15$\sigma$ error circles (=$[-2ln(1-0.9)]^{1/2}$=2.15), that is 90\% confidence radius as the possible optical counterparts (See Figure~\ref{fig7}.) While checking the archival HST data we also found bright sources lying within the 90$\%$ error circle of SN 2014C taken at t = 600 d (HST ObsID 14202) and t = 1017 d (HST ObsID 19668) with WFC3. The detailed analysis will be carried out in our later paper of SN 2014C analysis. 

\subsection{Optical counterparts identification} \label{sec:counterparts}

\begin{figure*}[htbp]
\figurenum{8}
\epsscale{0.2}
\centering
\includegraphics*[width=0.95\textwidth]{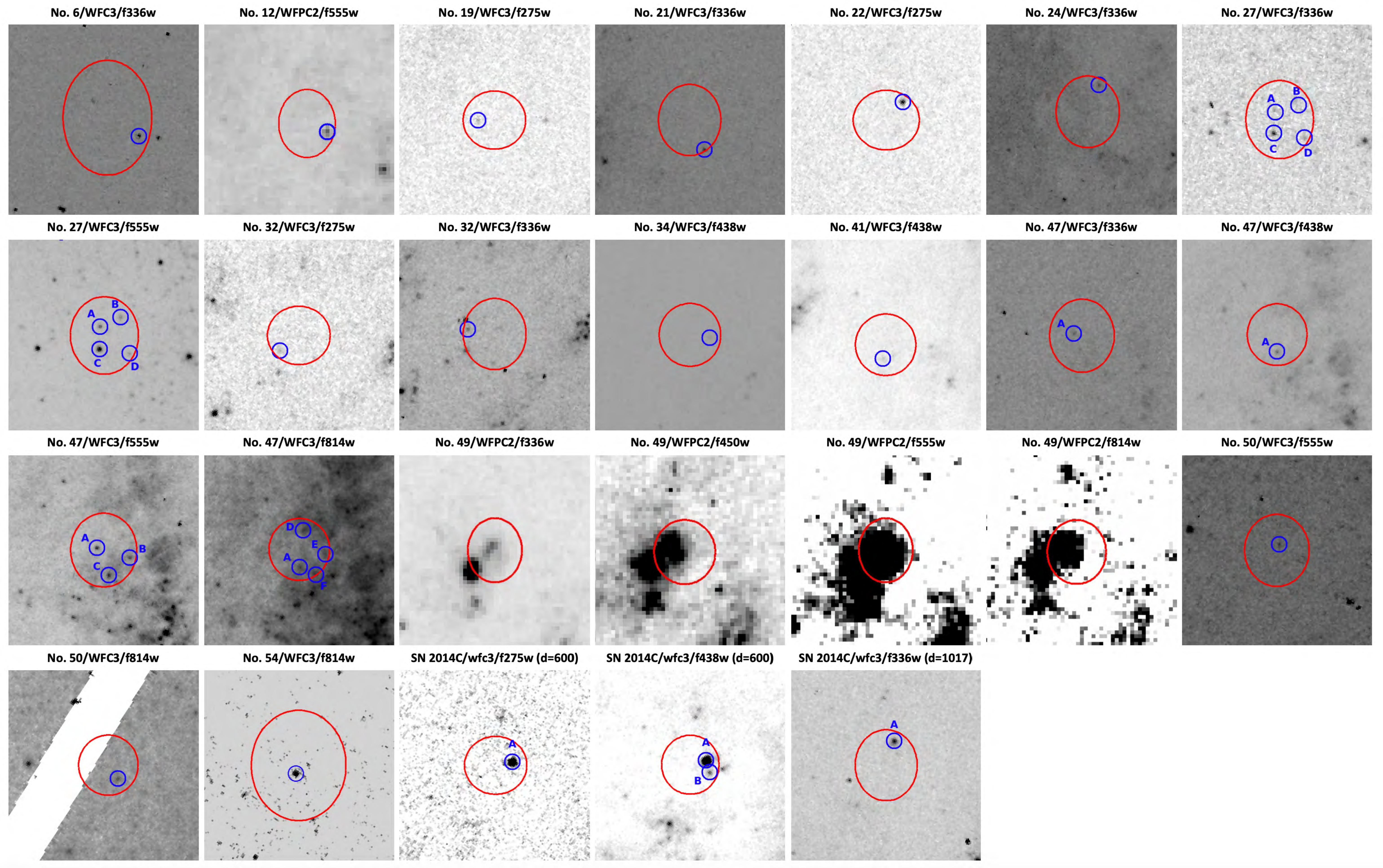}
\caption{Representative 5$\arcsec\times5\arcsec$ finding charts for the optical counterpart candidates with the X-ray 90\% confidence error ellipses overlaid in red. The optical counterpart candidates detected by \texttt{dolphot} are circled in blue.  For those containing multiple candidates are labeled with alphabet to distinguish them.}\label{fig7}
\end{figure*}

Possible optical counterparts are identified by overlaying the 90\% error circles of X-ray sources onto the HST images as shown in Figure~\ref{fig7} (red ellipses). Many of the detected X-ray point sources of NGC 7331 do not locate on any of the HST images or are within regions associated with dense clouds. Fourteen X-ray sources (excluding SN 2014C) are found to have detected optical sources locating within their error ellipses (marked with blue circles in Figure~\ref{fig7}). Sources found in different filters are checked with their relative positions to other bright sources in the field of view (FOV) of the image. If multiple sources are found in one error ellipse, they are labeled with letters to distinguish them. Since the optical counterpart of No. 49 is associated with a clump of unresolved sources instead of a point source as shown in Figure~\ref{fig7}, it will be discussed in details in Section~\ref{subsec:identification}.

As for the 21 candidates corresponding to the remaining 13 X-ray sources, we further calculate the ratio of X-ray to optical flux log(f$_{X}$/f$_{Opt}$) to roughly distinguish the optical counterparts of the X-ray sources in NGC 7331 from distant active galactic nuclei (AGN) by adopting the classification determined by the Chandra multiwavelength project (figure 5 in \cite{gre04}). The X-ray flux is calculated in the 0.5 -- 2 keV band. For the optical counterpart candidates detected in both B band (F438W) and V band (F555W) observations, we convert the B and V magnitude to SDSS r' magnitude according to the conversion formula, r' = 1.46V - 0.46B + 0.11 \citep{fuk96} and derive the ratio of optical to X-ray fluxes using Equation~\ref{eq:4} in \cite{gre04}:

\begin{equation}\label{eq:4}
{ log }_{ 10 }\frac { { f }_{ X } }{ { f }_{ r' } } = { log }_{ 10 }({ f }_{ X })+\frac { r' }{ 2.5 } +5.67. 
\end{equation}

Alternatively, we calculate the ratio of optical to X-ray fluxes for the sources with only V magnitudes by following \cite{lay17} as :

\begin{equation}\label{eq:5}
{ log }_{ 10 }\frac { { f }_{ X } }{ { f }_{ V } } = { log }_{ 10 }({ f }_{ X })+\frac { V }{ 2.5 } +5.37. 
\end{equation}

Seven of these 21 potential optical counterparts are detected in either both B band (F438W) and V band (F555W) or only in V band and they can be used for determining the flux ratio based on Equation~\ref{eq:4} and \ref{eq:5}.  We find that 5, No. 27 (candidate A, B and D), No. 47 (candidate A only), and No. 50, among these 7 optical sources are unlikely to be AGNs because their log(f$_{X}$/f$_{Opt}$) values are larger than one. Nevertheless, log(f$_{X}$/f$_{Opt}$) for most of the AGNs should lie in the range of -1 and 1 \citep{gre04, lay17}. Two of the five non-AGN candidates, No. 27 and No. 50, are also classified as LMXBs based on their X-ray colors.

Three optical counterpart candidates (No. 19, 22, and 32) are detected in WFC3 UV (F275W) bands while immersing in dense clouds in the longer wavelength HST observations. No. 19 and No. 32 are likely to be HMXBs according to the color-color classification (See Section~\ref{subsec:spectra}) and No. 32 is also defined as a variable (See Section~\ref{subsec:flux}). The converted absolute V  band magnitudes of the optical counterpart candidates of these two sources are about --10 to --9. It is suggestive that they may be OB stars, providing additional evidence that the two X-ray sources are HMXBs. However, we cannot rule out the chance that these two optical counterparts are composed by unresolved multiple sources because their absolute V magnitudes are much brighter than typical main sequence OB stars and their locations are very close to the spiral arm with dense clouds. 

However, it is possible that the above optical counterpart candidates are lying in the X-ray error circles by chance. To estimate the probabilities of positional coincidence of each source first, we follow \cite{lu_09} in calculating the Poisson probabilities of finding the number of observed optical sources within the X-ray error circle. There are two ways of estimating the average number of optical sources in the nearby region of the error circle. The first is taking the average optical sources found in 5$\arcsec\times$5$\arcsec$ finding charts centered at the X-ray error circle position to calculate the probability that we find the sources by just coincidence. Another is to find the average number of sources by shifting the 5$\arcsec\times$5$\arcsec$ finding charts 5$\arcsec$ from the X-ray error circle center to the north, east, south, and west. Unfortunately, the random probabilities for all the optical counterpart candidates are higher than 10$\%$ for both methods. That is partly due to the contamination of nearby clouds. 

\section{Discussion} \label{sec:discussion}

\subsection{X-ray population} \label{subsec:population}

By combining seven Chandra observations of NGC 7331 with a total exposure time of 126 ks, we detect 55 X-ray point sources with 5 estimated background AGNs in the field of our observations. The unabsorbed luminosities range from 7.0$\times10^{37}$ erg s$^{-1}$ to 1.6$\times10^{40}$ erg s$^{-1}$ in 0.3 -- 8 keV by assuming a power-law with $\Gamma$ = 1.7 for all the S/N $>$ 3 detected sources in the datasets. The lowest luminosity with S/N $>$ 2 detection limit is 3.6$\times10^{37}$ erg s$^{-1}$, which is comparable to Zezas et al. (2001). The maximum detected luminosities of 11 sources are larger than 1$\times10^{39}$ erg s$^{-1}$ and 2 of them are even more luminous than 2$\times10^{39}$ erg s$^{-1}$. We found 8 more ULXs than previous ULX study done by \cite{swa11}, which is based only on one observation (ObsID 2198).

In our study, the slopes ($\alpha_{1}$) of XLFs fitted with a simple power-law for individual observation range from 1.06 (HPD = 0.97 -- 1.15) (ObsID 02198) to 1.42 (HPD = 1.28 -- 1.53) (ObsID 17570) (See Section~\ref{tab:xlf_fitting}). If we fit with the pooled and hierarchical algorithm, the slopes for the power-law model are 1.18 (HPD = 1.12 -- 1.24) and 1.21 (HPD = 1.15 -- 1.22) respectively. Both of them lie in between the steepest and flattest mean slope based on individual method.  Moreover, the power-law with an exponential cut-off (PLC) model can also describe the cumulative XLFs with a slope ($\alpha_{2}$) of 0.62 (HPD = 0.49 -- 0.75) and a cut-off at around 1.05 (HPD = 0.86 -- 1.31)$\times$10$^{39}$ erg s$^{-1}$ with the pooled data in our study. If we adopt the same (PLC) model using the hierarchical method, the slope ($\alpha_{2}$) decreases to 0.48 (HPD = 0.36 -- 0.60) and a cut-off at around 0.8 (HPD = 0.58 -- 1.09)$\times$10$^{39}$ erg s$^{-1}$. However, the PLC model can only be constrained in the individual fitting for the first observation (ObsID 02198) with the means of the slope and the cut-off luminosity similar to the results in pooled algorithm. It is probably because the largest detected source number in ObsID 02198 or the PLC model is not a good model to describe the XLFs of other 6 datasets. This also demonstrates that the XLF in a pooled model is biased by the observation with more data points.

The early studies of Galactic as well as extragalactic XRBs reveal two distinct populations: the short-lived HMXBs and the long-lived LMXBs \citep{gri02}. Late-type star-forming galaxies dominated with HMXBs usually can be fitted well with a simple power-law XLF, while early-type galaxies more associated with LMXBs are best described with a broken power-law with a break at the Eddington luminosity caused by neutron star and/or black-hole XRB populations \citep{siv03, kon03, col04, zez02, kil02, kim10}. The broken-powerlaw XLF distribution suggests a combination of different X-ray populations. According to a study of XRBs in the Milky Way, the cumulative slope of XLF is $\sim$ 0.26 for Galactic LMXB with a luminosity cut at $\sim$ 2.7$\times$10$^{38}$ erg s$^{-1}$ and $\sim$ 0.64 for Galactic HMXB \citep{gri02}. Our PLC best fitted XLF slope ($\sim$ 0.5--0.6) lies between the two implying that the X-ray sources in NGC 7331 are likely a combination of LMXBs and HMXBs. Moreover, its cut-off at about 10$^{39}$ erg s$^{-1}$  (HPD = 0.86 -- 1.31 $\times$10$^{39}$ erg s$^{-1}$) is also higher than the truncated X-ray luminosity of the Galactic LMXB XLF. This may be due to the addition of HMXBs in the high end tail of the XLF. High luminosity cut-off at about few 10$^{39}$ erg s$^{-1}$ in our XLF analysis was also reported in the studies of NGC 4365, NGC 4382 \citep{siv03} and LMXBs in early type galaxies \citep{kim04, gil04}.

For nearby galaxies, \cite{col04} reported the slope of the cumulative PL XLF is $\sim$ 0.6--0.8 for spiral and starburst galaxies and is $\sim$ 1.4 for elliptical galaxies based on a sample of 32 galaxies. Our cumulative slope of $\sim$ 1.2 in a simple power-law model, no matter what the algorithm is, is also in between these two categories. The reason for the steeper slope of NGC 7331 than their spiral and starburst samples is very likely to be the contamination of the older X-ray sources located in the bulge due to the high inclination that the sources are less able to be spatially separated than the other face-on galaxies. Therefore, the X-ray sources include not only those younger population located on the spiral arms but also those in the bulge. Previous XLFs studies of face-on nearby galaxies confirm that the XLFs of older population often are steeper or more likely possess a break than those on younger stellar population (e.g, \cite{kon03}, \cite{swa03}, \cite{ten01}, \cite{sor02}.)

Some previous studies suggest that XLF of a galaxy might vary due to variability of X-ray sources \citep{bin17}, while others consider a ``snapshot" observation can still provide a good approximation to the XLF of a galaxy \citep{bin17, sel11}. Our findings in individual observation seem to support the former because the XLFs do change from observation to observation and it is very likely due to the impact of very bright outliers in XLFs. The bright end tail of the XLFs would bias the slope, especially for the cases with limited detected sources.

Nevertheless, this problem can be avoided by using the pooled or the hierarchical algorithm with Bayesian analysis to obtain a representative XLF. Therefore a series of monitoring over a galaxy is important in reducing the effect of outliers and providing a more reliable population study. For finding a representative XLF, the most intuitive way to infer its parameters is by pooling all the data into a single dataset. However, it would presumably be biased by the observation with the most detected sources as we see in the pooled PLC model fitting. In the PLC model, only the first observation (ObsID 02198) can be constrained in the individual fitting. Its means of the slope and the cut-off luminosity are close to the results in pooled algorithm. It is probably because of the largest detected sources number in ObsID 02198 observation or the PLC model is not a good model to describe the XLFs of other 6 datasets. This also demonstrates that the XLF in a pooled model could be biased by observations with more data points.

On the other hand,  this bias can be migrated by assuming that each individual data set is distributed according to a group distribution with the hierarchical method. Therefore, we can get a representative XLF out of the hierarchical model approach without ignoring the difference among different observations, such as the exposure time or the FOV. A hierarchical model can combine both group tendencies as well as individual differences \citep{ahn11} and it is widely proved to be able to provide reliable estimates compared to the separate estimation for individual data set \citep{kat16}. In particular, the shrinkage effect of a hierarchical model would move the parameter estimates toward the population mean as shown in Table~\ref{tab:xlf_fitting} by weighting. Estimates for observations with less detected sources tend to shrink towards the population mean while those for the datasets containing more sources will be closer to the average of the individual observation as stated in \cite{efr77} and \cite{gel06}.

\subsection{Source identification} \label{subsec:identification}

We crosscheck all the sources detected in this study with the Chandra Source Catalog (CSC Release 2) and XMM catalog (XMM-Newton Serendipitous Source Catalog (3XMM DR7 Version) produced by the XMM-Newton Survey Science Centre (SSC) consortium. Only 9 were in the XMM catalog but 39 coincide with CSC sources. Since CSC was based on the detection of the observation back in 2009, ObsID 2198, the 12 newly found X-ray sources in our study are either outside of the coverage of ObsID 02198 (2 sources), not detected (9 sources), or located in the crowded central region (1 source) which is resolved in the combined subpixel image in our study. 

Eight of the 55 detected sources can be classified to be possible X-ray binaries according to their X-ray colors and 13 are located in the overlapping classification regions. Six among these 8 candidates are similar to Galactic X-ray binaries showing bimodal high/soft and low/hard features. 

The combined spectra of the 24 non-variable bright sources in NGC 7331 apart from SN 2014C can be fitted by an absorbed power-law $\Gamma \simeq$ 1.7. (See Table~\ref{tab:spectral_fitting}.) This further supports the fact that many of them are very likely to be LMXBs. Among the bright X-ray sources, seven of them are located in the crowded central half D$_{25}$ region (labeled with ``c" in Table~\ref{tab:PointSource_table}) of NGC 7331, and we can not exclude the possibility that they are composed of several unresolved X-ray sources and are also contaminated by the diffuse X-ray emission. Forty-four sources are enclosed in the half $\mathrm{D_{25}}$ region of NGC 7331 but it is unclear if they are really in the bulge or just in the line of sight of the galactic central part. 

One of the ULXs in our study, source No. 49, is an absorbed source according to the X-ray color-color diagram classification (See Section~\ref{subsec:spectra}) and also a variable with the maximum to minimum flux ratio larger than 10. After it was discovered with Chandra, \cite{abo07} found that it is spatially associated with a bright star cluster (See Figure~\ref{fig7}) of mass M = 1.1 ($\pm$ 0.2)$\times$10$^{5}$ M$_{\sun}$. If source No. 49 is indeed part of this cluster, its progenitor of the compact accretor should have evolved through a supernova stage which implies its mass surpasses 40-50 M$_{\sun}$. To power the ULX, the companion star should be an evolved massive star in order to provide the high mass accretion supply. They also inferred that this ULX is the possible source for the shock excitation producing the observed [\ion{Si}{2}], [\ion{O}{1}], and [\ion{N}{2}] line intensities. 

We did not find any promising optical counterparts for the X-rays sources except for SN 2014C and No. 49 in the HST WFPC2 (F336W, F450W, F555W, and F814W) and WFC3 images (F275W, F336W, F438W, F555W, and F814W), but there are 21 optical counterpart candidates found lying within the other 13 X-ray sources error ellipses. Three X-ray sources (No. 27, 49 and 47) classified as LMXB candidates based on the X-ray colors (See Section~\ref{subsec:flux}) have optical counterpart candidates that are unlikely to be AGNs inferred by their X-ray to optical flux ratios \citep{gre04}. The optical counterpart candidates of the other two possible HMXBs (No. 19 and No. 32) could be OB stars according to their very bright converted absolute V magnitudes ($\sim$ --10 to --9), which further supports that they are HMXBs. However, it is more likely that they are associated with unresolved clusters because the V magnitudes are much brighter than the typical main sequence OB stars. Unfortunately, we cannot exclude the possibility of chance positional coincidence of any of the optical counterpart candidates owing to the contamination of nearby clouds. 

\subsection{SN 2014C} \label{subsec:sn2014c_2}

The first light of SN 2014C was estimated to be on 30 December 2013 $\pm$ 1 day (MJD 56656 $\pm$ 1) by modeling the barometric light-curve \citep{mar17}. A series of observations and studies in optical, X-ray and radio wavelengths have been launched after the explosion of SN 2014C \citep{mil15,mar17,bie18}. These observations reveal that SN 2014C is a core collapse type Ib SN at first and becomes a type IIn SN after its shock encountering the shell stripped from the progenitor's hydrogen envelope. The SN shock wave of SN 2014C is believed to begin interacting with the dense H-shell at an age between t $\sim$ 100 and 200 d \citep{mil15, mar17} and to become a type IIn supernova according to the light curve. 

SN 2014C has also been reported its interaction with a circumstellar shell based on the 8.4 and 22.1 GHz images of Very Long Baseline Interferometry (VLBI) taken during t = 384 and 1057 d after its explosion \citep{bie18}. Their study covering the last five Chandra observations time span of our study estimates an average expansion speed of 19 300 $\pm$ 790 km s$^{-1}$ at t = 384 d and it slowed down to or remained at 13 600 $\pm$ 640 km s$^{-1}$ between t = 384 and 1057 d. They pointed out that the expansion velocity dropped about 30 $\%$ after t = 384 d based on the VLBI measurements, and they suggested that the forward shock had already exited the dense CSM shell by t = 384 d.

The first detection by Chandra is taken after the explosion at t = 308 d (ObsID 16005), with an unabsorbed 0.3--8 keV luminosity of 1.05($_{-0.30}^{+0.08}$)$\times$10$^{40}$ erg s$^{-1}$ by assuming a power-law spectrum model. As shown in Figure~\ref{fig5}, its brightness keeps on rising and reaches to 1.49($_{-0.45}^{+0.08}$)$\times$10$^{40}$ erg s$^{-1}$ at t = 477 d and remains at constant afterward. Its X-ray luminosity is much higher than the upper limit (2$\times$10$^{37}$ erg s$^{-1}$) of another type II supernova, SN 2013bu, in NGC 7331. \cite{and17} also report the rebrightening of radio emission at t $\sim$ 400 d. After that, the radio fluxes decrease in the following two years (See Figure~\ref{fig5}). This is also reported in \cite{mar17} according to the inferred 0.3 -- 100 keV X-ray luminosity. In our analysis using both Chandra and NuSTAR data, the 0.3 -- 30 keV luminosity has also kept stable between the epoch of 2 years to 3 years after explosion (See Figure~\ref{fig5}.)

Based on our spectral analysis, the column density decreases as the shock propagating outward (See Figure~\ref{fig5} and Table~\ref{tab:SNspec}), that is also reported in \cite{mar17}. Their best-fit N$_{H}$ estimated by using an absorbed thermal Bremsstrahlung model are 2.9($^{+0.4}_{-0.3}$)$\times$10$^{22}$ cm$^{-2}$ and 1.8($^{+0.2}_{-0.2}$)$\times$10$^{22}$ cm$^{-2}$ for ObsID 17569 ( t = 397 d) and 17570 ( t = 477 d) respectively, which are close to our findings, 2.78($^{+0.72}_{-0.68}$)$\times$10$^{22}$ cm$^{-2}$ and 2.44($^{+0.52}_{-0.48}$)$\times$10$^{22}$ cm$^{-2}$. By including more later time data in this study, we further confirm that the absorption has continued to decrease after t = 477 d and lasted at least until t = 1029 d with a value of 0.61($^{+0.08}_{-0.08}$)$\times$10$^{22}$ cm$^{-2}$. As the column density decreases with time, the softer photons are detectable and the estimated absorbed fluxes in 0.3 -- 30 keV also rises a bit while the unabsorbed fluxes still remain constant (See Figure~\ref{fig5}). The estimated 0.3 -- 2 keV absorbed flux increases from $\sim$ 1.5 to 9.8$\times$10$^{-12}$ erg cm$^{-2}$ s$^{-1}$ during t = 397 to 1029 d.

The plasma temperatures inferred by an absorbed thermal Bremsstrahlung model in \cite{mar17} are 17.8($^{+3.7}_{-2.8}$) and 19.8 ($^{+6.3}_{-3.9}$) keV 1 year and 1.5 years after the explosion, respectively. These are slightly higher than our results by using a \texttt{vapec} model but are still in the same order. Our best fitted temperatures stay in $\sim$ 10 to 20 keV with only a slight decline although they are all within the uncertainties. At this temperature range, the most possible detectable X-ray line emission are emitted by the transitions related to the H- or He-like Fe atoms. These emission lines are apparent at all five epochs in our Chandra-NuSTAR datasets. However, a solar abundance Fe value fails in the spectral fitting by using the density of $\sim$ 10$^{6}$ cm$^{-3}$ and the temperatures of around 20 keV unless we use super solar abundance values. The estimated Fe abundances based on our \texttt{vapec} model are about $\sim$2 to 5 solar Fe abundances. Another possible interpretation of the excess Fe line  emission is a global asymmetry or the clumpy structure of the cooler or denser gas of CSM. The excess Fe emission spectral character has also been reported in other type IIn supernovae, such as SN 2006jd \citep{kat16, cha12} and SN 2009ip \citep{mar14}.

\section{Conclusion} \label{sec:conclusion}

In this paper, we report on the first study of the variability of X-ray sources in NGC 7331. We detect 55 X-ray sources with source significance $>$ 3 in the $\mathrm{D_{25}}$ region of NGC 7331 based on the Chandra ACIS-S observations with a total exposure time of 126 ks spanning over $\sim$ 16 years. The detection limit of our sample is 7.0$\times10^{37}$ erg s$^{-1}$ in the 0.3 -- 8.0 keV energy range. Thirteen sources are defined as variables based on the variability parameter. Most of the variables with counts $>$ 20 are likely to be XRBs according to the classification by their X-ray colors. Ten of the 55 X-ray sources possess the high/soft-low/hard bimodal feature, which is common for Galactic XRBs. The spectra of the brighter 24 sources can be fitted with a simple power-law with a photon index ranging from 0.4 to 2.4 with an average of 1.7. The fitted $\mathrm{N_{H}}$s are $\sim $10$^{21}$ cm$^{-2}$. Although we have found 21 optical counterparts candidates lying within the X-ray error ellipses and confirmed 7 of them are unlikely to be AGSs, we still can not rule out the possibility that it is due to positional coincidence.

As for the brightest source, SN 2014C, its spectrum (0.3 -- 30 keV) shows a very distinct emission peaking at $\sim$ 6.7 -- 6.8 keV which is associated with He-like or H-like Fe transitions. As the expansion of SN 2014C continues, the unabsrobed X-ray luminosity reaches a maximum at 3.5($\pm$ 0.7)$\times$10$^{40}$ erg s$^{-1}$ by 400 days after the explosion and stays almost constant in the following two years in the 0.3 -- 30 keV energy range. The decrease of column density causes the rising of detected soft photons during the second and third year after its explosion. Apart from a power-law with a Gaussian profile model, SN 2014C can also be fitted by a collisionally-ionized ionization equilibrium thermal model. It suggests the plasma temperatures are around 10 to 20 keV and the luminous Fe emission line probably originates from the asymmetry or the clumpy structure of the CSM of SN 2014C.
 
More than one third of the X-ray sources are variables in NGC 7331 and the XLF at different epoch in our study does differ much with each other according to our XLF analysis. This conclusion is inconsistent with the early suggestion that a ``snapshot" observation can still provide a good approximation to the XLF of a galaxy \citep{bin17, sel11}. We also show that fitting the XLF based on multiple observations can largely eliminate the impact of the variable luminous end and provides a more reliable population statistics of a galaxy. Our study in XLF analysis is also the very first one adopting Bayesian algorithm. This approach provides a more robust statistics in the cases with a low source count such as NGC 7331. We also introduce a hierarchical approach, which has been already widely adopted in other fields  such as sociology and biology, for obtaining the representative XLF of NGC 7331 without losing the information at different epochs. Our result shows that a power-law with an exponential cut-off using a hierarchical fitting can best describe the XLF of NGC 7331 suggesting a combination of LMXBs and HMXBs in the X-ray population.

\bigskip
We thank H.F. Yu for useful discussions of Bayesian analysis. This project is supported by the Ministry of Science and Technology of the Republic of China (Taiwan) through grants 105-2119-M-007-028-MY3 and 107-2628-M-007-003.



\end{document}